\documentclass[aps,PRR,twocolumn,superscriptaddress,longbibliography]{revtex4-1}
\usepackage[colorlinks=true, allcolors=blue]{hyperref}
\usepackage{upgreek} 
\usepackage{natbib}
\usepackage{graphicx}
\usepackage{ulem}
\usepackage{comment,xcolor}

\begin{document}

\title{Reconciling mean-squared radius differences in the silver chain through improved measurement and {\it ab initio} calculations}

\author{B. Ohayon}\email{bohayon@technion.ac.il}
\affiliation{
The Helen Diller Quantum Center, Department of Physics,
Technion-Israel Institute of Technology, Haifa, 3200003, Israel
}

\author{J. E. Padilla-Castillo}
\affiliation{Fritz-Haber-Institut der Max-Planck-Gesellschaft, Faradayweg 4-6, 14195 Berlin, Germany}

\author{S. C. Wright}
\email[]{sidwright@fhi-berlin.mpg.de}
\affiliation{Fritz-Haber-Institut der Max-Planck-Gesellschaft, Faradayweg 4-6, 14195 Berlin, Germany}

\author{G. Meijer}
\affiliation{Fritz-Haber-Institut der Max-Planck-Gesellschaft, Faradayweg 4-6, 14195 Berlin, Germany}

\author{B. K. Sahoo}\email{bijaya@prl.res.in}
 \affiliation{
Atomic, Molecular and Optical Physics Division, Physical Research Laboratory, Navrangpura, Ahmedabad 380058, Gujarat, India
}

\date{\today}

\begin{abstract}
Nuclear charge radius differences in the silver isotopic chain have been reported through different combinations of experiment and theory, exhibiting a tension of two combined standard errors.
This study investigates this issue by combining high-accuracy calculations for six low-lying states of atomic silver with an improved measurement of the $5s\, ^2S_{1/2} - 5p\, ^2P_{3/2}$ transition optical isotope shift.
Our calculations predict measured electronic transition energies in Ag I at the 0.3\% level, the highest accuracy achieved in this system so far. We calculate electronic isotope shift factors by employing analytical response relativistic coupled-cluster theory, and find that a consistent charge radius difference between $^{107,109}$Ag is returned when combining our calculations with the available optical isotope shift measurements. We therefore recommend an improved value for the mean-squared charge radius difference between $^{107}$Ag and $^{109}$Ag as $0.207(6)$ fm$^2$, within one combined error from the value derived from muonic Ag experiments, and an updated set of charge radii differences across the isotopic chain.
\end{abstract}

\maketitle

\section{Introduction}

The mean-squared nuclear charge radius difference between isotopes with nuclear mass numbers $A$ and $A'$, $\delta r^2_{A,A'}\equiv r^2_{A'}-r^2_{A}$, is a unique probe of structural changes in isotopic chains~\cite{2023-Review}, complementary to the binding energy per nucleon. As described in Ref.~\cite{2004-FH}, $\delta r^2_{A,A'}$ values can be inferred from measured isotope shifts (ISs) $\delta \nu_{A,A'}\equiv \nu_{A'}-\nu_A$ using the relation
\begin{eqnarray}
\delta \nu_{A,A'} & \simeq & 
K\mu_{A,A'} + F \delta r^2_{A,A'},
\label{eq:IS}
\end{eqnarray}
where $\mu_{A,A'}=1/M_{A}-1/M_{A'}$ is the difference between inverse nuclear masses of isotopes, $K$ denotes the mass shift (MS) factor, and $F$ the field shift (FS) factor of a given transition with frequency $\nu$.
The validity and refinements to Eq.~(\ref{eq:IS}) are discussed in section~\ref{sec:val}.

When $\delta r^2_{A,A'}$ of two or more isotopic pairs have been measured, usually via muonic atom cascade X-ray spectroscopy~\cite{1969-Barret}, the atomic factors $K$ and $F$ of Eq.~(\ref{eq:IS}) can be directly extracted from a linear fit called a calibrated King Plot (CKP)~\cite{1963-KP}, 
having two or more data-points.
Using this information, $\delta r^2_{A,A'}$ can then be extracted across a chain of isotopes via optical isotope shift measurements, and without further muonic atom experiments. This is the case for most elements with an even number of protons ($Z$)~\cite{2004-FH}.
For the odd-$Z$ elements,
there aren't three or more stable isotopes available that are needed for carrying out traditional cascade spectroscopy measurements (see, however, Ref. \cite{2023-Micro,2023-Implanted}). So in order to apply Eq.~(\ref{eq:IS}) to extract $\delta r^2$ values in a chain of isotopes, one has to rely on the calculation of the IS factors $F$ and $K$.

For elements in which $\delta r^2_{A,A'}$ for a pair of isotopes has been determined, it is sufficient to calculate or estimate one of the two IS factors, and extract the other factor via Eq.~(\ref{eq:IS}).
It is convenient to calculate the $F$ values as they are less susceptible to electron correlation effects as compared to $K$.
This method is sometimes referred to as a partial CKP. It is used in odd-$Z$ elements with at least two stable isotopes (see e.g. \cite{2016-Cu, 2012-Ga, 2015-Rb}).
It is also useful for light even-$Z$ elements in which a CKP yields larger uncertainties (e.g.~\cite{2004-Ti,2008-Ar,2019-Ne,2022-Na, 2023-Si}).
The partial CKP method has been considered to give both precise and accurate results for $\delta r^2_{A,A'}$, as it relies on a calculation of $F$, much easier than of $K$, and the use of $\delta r^2$ from muonic cascade measurements which are considered reliable at a few attometer level.
Nevertheless, with tremendous advancement in the development of atomic many-body methods, as well as availability of ever-increasing computational power, it is possible today to calculate the $K$ values precisely in some systems.
This enables extracting \textit{all optical} $\delta r^2_{A,A'}$ beyond the accuracy obtained in either a partial or a full CKP approach~\cite{2020-In,2021-K,2021-Yb,2022-Cd,2022-CdHG,2023-Zn,2023-In,2023-Tl,2024-Al}, which are forever limited by complex nuclear corrections to the muonic energy levels.

\begin{figure*}[!t]
\centering
\includegraphics[trim=4 96 20 126,clip,width=0.98\textwidth]{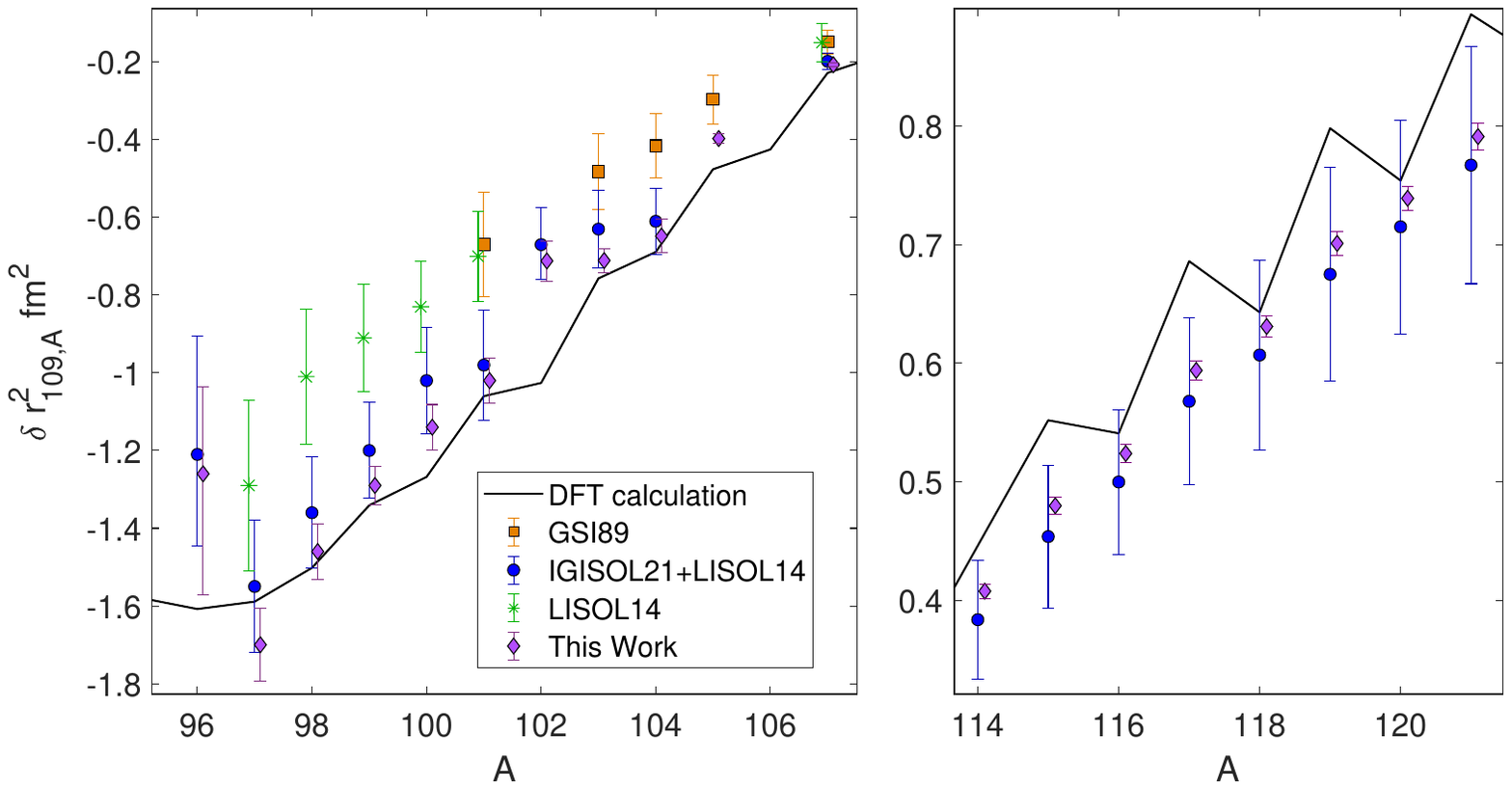}
\caption{Mean-squared charge radius difference, $\delta r^2_{109,A}$, of ground-state silver nuclei as given in Table~\ref{tab:rad}. Error bars show $68\%$ confidence intervals taking into account both statistical and systematic uncertainties.
The results of a density-functional-theory (DFT) calculation from Ref.~\cite{2021-Ag} are also shown.
The results of this work are in disagreement with the GSI~\cite{1989-GSI} and LISOL~\cite{2014-LISOL} measurements and analysis, and agree with recent work in IGISOL~\cite{2021-Ag}, though with significantly reduced uncertainties.
}
    \label{fig:rad}
\end{figure*}
A particularly interesting case is that of silver (Ag, $Z=47$), having two stable isotopes with a similar natural abundance, $^{107}$Ag and $^{109}$Ag. 
ISs of three transitions in Ag with unstable nuclei have been measured.
%
In 1975, the $5s$$~^2S_{1/2}$$-5p~^2P_{1/2}$ $338\,$nm and the $5s$$~^2S_{1/2}$$-5p~^2P_{3/2}$ $328\,$nm lines were measured with a hollow-cathode Fabry–P\'erot interferometer using neutron-irradiated targets of $^{108m}$Ag and $^{110m}$Ag~\cite{1975-108}.
%
The first online experiment was performed at the Gesellschaft f{\"u}r Schwerionenforschung (GSI) accelerator facility, where ISs in the $4d^95s^2~^2D_{5/2}-6p~^2P_{3/2}$ $548\,$nm line were measured for neutron-deficient isotopes and isomers~\cite{1989-GSI}.
Further neutron-deficient isotopes were measured with the $328\,$nm line at the Leuven Isotope Separator On Line (LISOL) facility~\cite{2014-LISOL}.
Recently, at the Ion Guide Isotope Separation On-Line (IGISOL) facility, ISs of a long chain of isotopes, extending from $^{96}$Ag to $^{121}$Ag were measured; again using the $328\,$nm line~\cite{2021-Ag}.
The center-of-gravity ISs resulting from these efforts have been interpreted via partial CKPs, with different choices for $\delta r^2_{109,107}$ and $F$, yielding inconsistent results for $\delta r^2_{109,A}$, as shown in Fig.~\ref{fig:rad}.

In this work we resolve the tension in $\delta r^2$ of Ag reported thus far.
To do this, we have performed state-of-the-art \textit{ab initio} calculations of IS factors in low-lying levels of Ag~I.
Our results indicate that semi-empirical estimations of $K$ and $F$, used previous studies of Ag, deviate by two combined standard errors.
We also provide improved optical isotope shift of the $328\,$nm transition in naturally abundant Ag, and perform a global analysis of the result with the available literature data.
Our calculations of $K$ and $F$ for these transitions in Ag, combined with the data, produce a consistent value for $\delta r^2_{109,107}$ within a few percent.
This checks the consistency of our calculations. 

Having validated our calculation, we use the available optical IS data to provide $\delta r^2_{109,A}$ for the silver isotope chain spanning $A=96-105$, $A=114-121$, and six long-lived isomers. These results reduce the uncertainty in $\delta r^2_{109,A}$ by up to a factor of seven.
A comparison with prior works pinpoints the reasons for past inconsistencies.
Finally, the recommended $\delta r^2_{109,A}$ of the Ag isotopes are compared with state-of-the-art nuclear theory calculations. 
Whilst the overall trend of the nuclear calculations agrees with the data extracted by combining our atomic calculations with the IS measurements, a few interesting deviations are noticed.

\section{Validity of linear approximation}\label{sec:val}

Before calculating the IS factors, we discuss refinements to Eq.~(\ref{eq:IS}), in light of our aim of extracting $\delta r^2$ in a long chain of isotopes, in which the charge radius is expected to vary significantly.
As seen in Fig.~\ref{fig:rad}, the isotopic difference in the mean-squared radius can be as large as $2.6\,$fm$^2$, so that the root-mean-squared radii span the range $R=4.37-4.64\,$fm.
In this work, we take into account a possible dependency of $K$ and $F$ on $R$ by repeating their calculation for different values of $R$ in a range larger than that spanned by the experimental values.

Another refinement of Eq.~(\ref{eq:IS}) is due to variation of the nuclear shape among the isotopes. The effect on the IS can be estimated via a change to the FS factor $\delta F_{A,A'}= C(rcc_{A'}^4/r_{A'}^2-rcc_{A}^4/r_{A}^2)$, with $rcc$ the fourth radial moment and $C\simeq -6 \times 10^{-4}~$fm$^{-2}$; the Seltzer coefficient calculated in~\cite{1987-Blund}.
As the ratio of moments depends on the nuclear shape, we estimate their maximal change by varying the skin thickness parameter by $10\%$ in Eq.~(\ref{eq:Fermi}), as suggested in Ref.~\cite{2004-FH}. 
It shows a 2\% variation in $\delta F_{A,A'}$, which is in agreement with the finding of Ref.~\cite{1989-GSI} (see their Eq.~(20)) that was also adopted by Ref.~\cite{2021-Ag}.
To be more conservative, and following the discussion given in Ref.~\cite{2023-Zn}, we treat this contribution not as a correction but as another source of uncorrelated uncertainty to the nuclear radii extraction in the Ag isotopes.

It is also worth noting that in this work ISs are defined as the centers-of-gravity of the hyperfine manifold. 
When there are nearby states of equal parity, the mixing of fine and hyperfine structure introduces a nuclear spin-dependant shift of the ISs (see e.g.~\cite{1991-HFS2}).
For the energy levels of Ag that are of interest to this work, the largest shift would be due to the mixing of fine and hyperfine structure in the $5p$ manifold.
The order of magnitude of this shift can be roughly estimated by $\Delta E^{(2)}\simeq A(5p~^2P_{1/2})A(5p~^2P_{3/2})/\Delta E_\mathrm{FS}$ \cite{Sahoo2003,Sahoo2015}, where $\Delta E_\mathrm{FS}=28~$THz is the fine structure of the $5p$ manifold. 
Even for the isotopes with the largest magnetic moments, $\Delta E^{(2)}\simeq~10~$kHz.
It is three orders smaller than the precision of interest to the present work, hence it can be safely neglected.

\section{Atomic Calculations}

\subsection{Method of calculation}

\begin{table*}[tb]
\caption{
Comparison of calculated and measured energies.
Calculated electron affinities (EAs, where zero energy refers to the ground state of Ag~II) of the considered states in Ag at different levels of approximation. The estimated excitation energies (EEs) are also quoted.
Unless otherwise stated, all values are in cm$^{-1}$.
Our final results are compared with the experimental values (denoted Exp.).
Differences between our calculated and experimental values are shown as $\Delta$ in percentage. The last column gives the results of prior calculations available in the literature which are closest to experiment.
}
\begin{ruledtabular}
\begin{tabular}{l rrrr rrrr rrr}
State  & \multicolumn{1}{c}{DHF}  & \multicolumn{1}{c}{MP2}  & \multicolumn{1}{c}{RCCSD} & \multicolumn{1}{c}{+T} & $+$Basis & $+$Breit & $+$QED  & \multicolumn{1}{c}{Total} & \multicolumn{1}{c}{Exp.~\cite{1998-IP,2001-absolute}} & \multicolumn{1}{c}{$\Delta$\%} & \multicolumn{1}{c}{Lit.}\\
\hline \\
EAs\\  
$5s$ $~^2S_{1/2}$ & $50376$ & $61014$ & $60408$ & $441(110)$ & $193(96)$ & $-59$ & $-22(22)$ & $60961(148)$ & $61106.5(2)$ & $0.2(2) $ & $60823$~\cite{2006-theory}\\
$5p~^2P_{1/2}$ & $26730$ & $30771$ & $31007$ & $455(114)$ & $46(23)$  & $-36$ & $4(4)$   & $31477(116)$  & $31554.4(2)$ & $0.2(4) $ & $31066$~\cite{2006-theory}\\
$5p~^2P_{3/2}$ & $26148$ & $29862$ & $30089$ & $442(111)$ & $39(20)$  & $-24$ & $-4(4)$  & $30543(112)$  & $30633.7(2)$ & $0.3(4) $ & $30184$~\cite{2006-theory}\\
$6s$$~^2S_{1/2}$ & $17115$ & $18641$ & $18455$ & $45(11)$ & $21(11)$ & $-8$ & $-3(3)$ & $18510(16)$        & $18550.3(2)$ & $0.2(1) $ & $18494$~\cite{2006-theory}\\
$6p~^2P_{1/2}$ & $11786$ & $12726$ & $12680$ & $89(22)$ & $9(5)$ & $-8$ & $1(1)$ & $12771(23)$           & $12809.0(2)$ & $0.3(2) $ & $12656$~\cite{2022-The}\\
$6p~^2P_{3/2}$ & $11618$ & $12502$ & $12467$ & $94(24)$ & $8(4)$ & $-5$ & $-1(1)$ & $12563(24)$          & $12606.6(2)$ & $0.3(2) $ & $12452$~\cite{2022-The}\vspace{1 mm}\\
EEs\\
$5s$$~^2S_{1/2}$$-5p~^2P_{1/2}$ & $23646$ & $30243$ & $29400$ & $-14(4)$  & $147(73)$ & $-24$ & $-25(25)$ & $29484(78)$  & $29552.061(1)$ & $0.2(3) $ & $29496$~\cite{2022-The}\\
$5s$$~^2S_{1/2}$$-5p~^2P_{3/2}$ & $24228$ & $31152$ & $30319$ & $-2(0.)$  & $154(77)$ & $-36$ & $-18(18)$ & $30418(79)$  & $30472.703(1)$ & $0.2(3) $ & $30451$~\cite{2021-Flam}\\
$5s$$~^2S_{1/2}$$-6s$$~^2S_{1/2}$ & $33261$ & $42373$ & $41953$ & $396(99)$ & $172(86)$ & $-51$ & $-19(19)$ & $42451(105)$ & $42556.152(2)$ & $0.2(3) $ & $42329$~\cite{2006-theory}\\
$5s$$~^2S_{1/2}$$-6p~^2P_{1/2}$ & $38590$ & $48288$ & $47727$ & $352(88)$ & $185(92)$ & $-51$ & $-23(23)$ & $48190(108)$ & $48297.402(3)$ & $0.2(3) $ & $47765$~\cite{2022-The}\\
$5s$$~^2S_{1/2}$$-6p~^2P_{3/2}$ & $38758$ & $48512$ & $47940$ & $347(87)$ & $185(93)$ & $-54$ & $-21(21)$ & $48398(103)$ & $48500.805(2)$ & $0.2(3) $ & $47969$~\cite{2022-The}\\
\end{tabular}
\end{ruledtabular}
\label{tab:E}
\end{table*}

We consider the Dirac-Coulomb (DC) Hamiltonian to calculate the IS factors in the relativistic framework, given in atomic units (a.u.) by
\begin{eqnarray}\label{eq:DC}
H &=& \sum_i \left [c {\vec \alpha}_i^D \cdot {\vec p}_i+(\beta_i^D-1)c^2+V_n(r_i)\right] +\sum_{i,j>i}\frac{1}{r_{ij}}, \ \ \ \
\end{eqnarray}
where $c$ is speed of light, ${\vec \alpha}^D$ and $\beta^D$ are the Dirac matrices, ${\vec p}$ is the single particle momentum operator, $V_n(r)$ denotes nuclear potential seen by an electron at distance $r$ from the nucleus and $\frac{1}{r_{ij}}$ represents the Coulomb potential between the electrons located at the $i^{th}$ and $j^{th}$ positions. The finite-size of the nucleus is defined by a two-parameter Fermi-charge distribution, given by
\begin{eqnarray}\label{eq:Fermi}
\rho(r) = \frac{\rho_0}{1+e^{(r-b)/a}} ,
\end{eqnarray}
where $\rho_0$ is the normalization constant, $b$ is the half-charge radius and $a=2.3/4\ln(3)$ is an approximate skin thickness.
It yields
\begin{eqnarray}
 V_n(r) = -\frac{Z}{\mathcal{N}r} \times \ \ \ \ \ \ \ \ \ \ \ \ \ \ \ \ \ \ \ \ \ \ \ \ \ \ \ \ \ \ \ \ \ \ \ \ \ \ \ \  \nonumber\\
\left\{\begin{array}{rl}
\frac{1}{b}(\frac{3}{2}+\frac{a^2\pi^2}{2b^2}-\frac{r^2}{2b^2}+\frac{3a^2}{b^2}P_2^+\frac{6a^3}{b^2r}(S_3-P_3^+)) & \mbox{for $r_i \leq b$}\\
\frac{1}{r_i}(1+\frac{a62\pi^2}{b^2}-\frac{3a^2r}{c^3}P_2^-+\frac{6a^3}{b^3}(S_3-P_3^-))                           & \mbox{for $r_i >b$} ,
\end{array}\right.   
\label{eq12}
\end{eqnarray}
where the factors are 
\begin{eqnarray}
\mathcal{N} &=& 1+ \frac{a^2\pi^2}{b^2} + \frac{6a^3}{b^3}S_3  \nonumber \\
\textrm{with} \ \ \ \ S_k &=& \sum_{l=1}^{\infty} \frac{(-1)^{l-1}}{l^k}e^{-lb/a} \ \ \  \nonumber \\
\textrm{and} \ \ \ \ P_k^{\pm} &=& \sum_{l=1}^{\infty} \frac{(-1)^{l-1}}{l^k}e^{\pm l(r-b)/a} . 
\end{eqnarray}
In the above expression, $b$ is obtained using the relation 
\begin{eqnarray}
    b\simeq\sqrt{\frac{5}{3} R^2 - \frac{7}{3} a^2 \pi^2} ,
\end{eqnarray}
where $R$ is the approximate root mean-squared radius and calculated from the empirical relation~\cite{1985-Emp}
\begin{equation} 
\label{eqR}
R \simeq (0.836 A^{1/3} + 0.57) ~ \textrm{fm}.
\end{equation}
We later show that the excitation energies and IS factors depend very weakly on the assumed value of the charge radius, thus validating the above approximations.
The FS operator, $F$, is defined in this case as
\begin{eqnarray}
 {\hat F} = - \frac{\delta V_n(r)}{ \delta R} = 
 - \frac{\partial V_n(r)}{ \partial b} \frac{\delta b}{ \delta R} 
 .
\end{eqnarray}

In the relativistic formulation, the NMS and SMS operators are given by
\begin{eqnarray}
O^{\textrm{NMS}} &=& \frac{1}{2}\sum_i \left ({\vec p}_i^{~2} - \frac{\alpha_e Z}{r_i} {\vec \alpha}_i^D \cdot {\vec p}_i \right. \nonumber \\ && \left. - \frac{\alpha_e Z}{r_i} ({\vec \alpha}_i^D \cdot {\vec C}_i^1){\vec C}_i^1 \cdot {\vec p}_i \right ),
\label{nmsexp}
\end{eqnarray}
and
\begin{eqnarray}
O^{\textrm{SMS}} &=& \frac{1}{2} \sum_{i\ne j} \left ({\vec p}_i \cdot {\vec p}_j - \frac{\alpha_e Z}{r_i} {\vec \alpha}_i^D \cdot {\vec p}_j \right. \nonumber \\ && \left. - \frac{\alpha_e Z}{r_i} ({\vec \alpha}_i^D \cdot {\vec C}_i^1) ({\vec p}_j \cdot {\vec C}_j^1) \right ),
\end{eqnarray}
respectively, where $\alpha_e$ is the fine-structure constant and ${\vec C}$ is the Racah angular momentum operator. Contributions from the Breit and lower-order QED interactions are also estimated, by including them self-consistently in the calculations~\cite{2016-bijaya1}, as corrections to the DC Hamiltonian results.

Expectation values of $F$, $O^{\textrm{NMS}}$ and $O^{\textrm{NMS}}$ with respect to wave function of an atomic state will correspond to the FS, NMS and SMS factors respectively. It should be noted from Eq.~(\ref{eqR}) that different $R$ values can affect $V_n(r)$ and hence, calculation of the atomic wave functions. We later validate that such changes in wave functions does not affect the energies, NMS factors and SMS factors at the precision of our interest, but it may affect calculations of the FS factors as the $F$ operator can have non-linear dependency on $R$. We demonstrate this dependency by calculating the energies and IS factors later with different values of $R$.

The Ag atom has the ground state configuration $[4d^{10}]~5s$. Our interest is to calculate the difference in the values between two states of the IS factors of the $5s$$~^2S_{1/2}$$-5p~^2P_{1/2;3/2}$, $5s$$~^2S_{1/2}$$-6s$$~^2S_{1/2}$ and $5s$$~^2S_{1/2}$$-6p~^2P_{1/2;3/2}$ transitions in this work. We calculate these factors for each state using the relativistic coupled-cluster (RCC) theory by adopting the analytical-response approach (AR-RCC method) as described in Ref.~\cite{2020-In}. It requires 
determination of wave functions of the ground state as well as for the $5p~^2P_{1/2;3/2}$, $6s$$~^2S_{1/2}$ and $6p~^2P_{1/2;3/2}$ states of Ag I. To obtain all these states conveniently, we first calculate wave function ($|\Psi_0 \rangle$) of the common closed-shell core configuration $[4d^{10}]$ of all these states by expressing it in the RCC theory ansatz~\cite{1980-CC}
\begin{eqnarray}
    |\Psi_0 \rangle = e^{S_0} |\Phi_0 \rangle ,
\end{eqnarray}
where $S_0$ is the RCC excitation operator carrying electron correlation effects and the reference state $|\Phi_0 \rangle$ is the mean-field wave function of the closed-shell $[4d^{10}]$ configuration, obtained in the Dirac-Hartree-Fock (DHF) method. The amplitude determining equation for $S_0$ is given by
\begin{eqnarray}
\langle \Phi_0^* |  \left ( H e^{S_0} \right )_l | \Phi_0 \rangle = 0 ,
\label{eqs0}
\end{eqnarray}
where $| \Phi_0^* \rangle $ represents all possible excited state determinants with respect to $| \Phi_0 \rangle$ and subscript $l$ means linked terms. First, we approximate the RCC theory at the singles and doubles approximation (RCCSD method), in which the $S_0$ is defined as
\begin{eqnarray}
S_0 &=& S_{10} + S_{20} \nonumber \\
    &=& \sum_{a,p} s_{ap} a_p^{\dagger} a_a + \frac{1}{4} \sum_{a,b,p,q} s_{ap,bq} a_p^{\dagger} a_q^{\dagger} a_b a_a ,
\end{eqnarray}
where $S_{10}$ and $S_{20}$ stand for single and double excitations of the RCC operator $S_0$ with the amplitudes $s_{ap}$ and $s_{ap,bq}$, respectively. Here $a,b$ denote for occupied orbitals and $p,q$ represent for unoccupied (virtual) orbitals. Following Eq.~(\ref{eqs0}), it yields 
\begin{eqnarray}
&& s_{ap} =  \frac{ \langle \Phi_a^p | H + [(He^{S_0})_l-H)]_{ofd} | \Phi_0 \rangle } {\epsilon_a - \epsilon_p } \nonumber \\
\textrm{and} && \nonumber \\
&& s_{ap,bq} =  \frac{ \langle \Phi_{ab}^{pq} | H + [(He^{S_0})_l-H)]_{ofd} |\Phi_0 \rangle } {\epsilon_a +\epsilon_b - \epsilon_p -\epsilon_q} ,
\end{eqnarray} 
where  $| \Phi_0^* \rangle $ are taken as $|\Phi_a^p \rangle = a_p^{\dagger} a_a |\Phi_0 \rangle $ and $|\Phi_{ab}^{pq} \rangle = a_p^{\dagger} a_q^{\dagger} a_b a_a |\Phi_0 \rangle $ representing the singly and doubly excited Slater determinants, respectively. Here the subscript $ofd$ denotes off-diagonal terms and $\epsilon$s are the single particle orbital energies.

After obtaining the solution for $|\Psi_0 \rangle$, we determine the wave function of an atomic state  ($|\Psi_v \rangle$) of Ag I with a valence orbital $v$ by defining~\cite{1983-LM,1989-MP}
\begin{eqnarray}
    |\Psi_v \rangle &=& e^{S_0+ S_v} |\Phi_v \rangle \nonumber \\
    &=& e^{S_0} \left \{ 1+S_v \right \} |\Phi_v \rangle ,
\end{eqnarray}
where $|\Phi_v \rangle = a_v^{\dagger} |\Phi_0 \rangle$ is the modified DHF wave function, and $S_v$ includes excitation configurations due to correlation effects by the valence electron. In the RCCSD method, we define
\begin{eqnarray}
S_v &=& S_{1v} + S_{2v} \nonumber \\
    &=& \sum_{p\ne v} s_{vp} a_p^{\dagger} a_v + \frac{1}{2} \sum_{p\ne v,b,q} s_{vp,bq} a_p^{\dagger} a_q^{\dagger} a_b a_v ,
\end{eqnarray}
where $S_{1v}$ and $S_{2v}$ stand for single and double excitations of the RCC operator $S_v$ with the amplitudes $s_{vp}$ and $s_{vp,bq}$, respectively. Like the case for the $S_0$ operator, amplitudes of the $S_v$ operator are determined by
\begin{eqnarray}
 \langle \Phi_v^* | \left \{ \left ( H e^{S_0} \right )_l -E_v) \right \} S_v  + \left ( H e^{S_0} \right )_l | \Phi_v \rangle = 0 , \label{eqamp}
\end{eqnarray}
where $| \Phi_v^* \rangle $ denotes for singly and doubly excited Slater determinants with respect to $| \Phi_v \rangle$. It corresponds to
\begin{eqnarray}
&& s_{vp} =  \frac{ \langle \Phi_v^p | (He^{S_0})_l + [(He^{S_0})_lS_v]_{ofd} | \Phi_v \rangle } {E_v - \epsilon_p } \nonumber \\
\textrm{and} && \nonumber \\
&& s_{vp,bq} =  \frac{ \langle \Phi_{vb}^{pq} | (He^{S_0})_l + [(He^{S_0})_lS_v]_{ofd} |\Phi_v \rangle } {E_v +\epsilon_b - \epsilon_p -\epsilon_q} . \ \ \ \ \
\end{eqnarray} 
The energy of the respective state is given by
\begin{eqnarray}
 E_v = \langle \Phi_v | \left ( H e^{S_0} \right )_l \{ 1+ S_v \} | \Phi_v \rangle . \label{eqeng}
\end{eqnarray}
Both Eqs.~(\ref{eqamp}) and~(\ref{eqeng}) are solved simultaneously by adopting a self-consistent procedure. We use here a normal-ordered Hamiltonian with respect to $|\Phi_0 \rangle$, so that the calculated $E_v$ value will correspond to the electron affinity (EA) rather than the total energy of $|\Psi_v \rangle$. Excitation energy (EE) of a transition can be obtained from the difference between EAs of two states associated with the transition.

In the AR-RCC method, we estimate IS factors as the first-order energy correction to the calculated $E_v$ value of the state $|\Psi_v \rangle$ due to the corresponding IS operator (denoted by $H_{\mathrm{IS}}$ in general). Hereafter, we identify the RCC operators and calculated energies due to the DC Hamiltonian with superscript $(0)$ and the first-order corrections in the AR-RCC method are denoted by the superscript $(1)$ as described in Ref.~\cite{2020-In}. In the AR-RCC method, the first-order energy correction ($E_v^{(1)}$) is obtained as the solution of the following equation  
\begin{eqnarray}
  (H-E_v^{(0)}) |\Psi_v^{(1)} \rangle &=& (E_v^{(1)} -  H_\mathrm{IS} ) |\Psi_v^{(0)} \rangle .
\end{eqnarray}
In the singles and doubles excitation approximation of the AR-RCC approach (AR-RCCSD method), the amplitudes of the $S_0^{(1)}$ and $S_v^{(1)}$ operators are defined in the similar way as in the case of unperturbed case and they are obtained by
\begin{eqnarray}
&& \langle \Phi_0^* |  \left ( H e^{S_0^{(0)}} S_0^{(1)} + H_\mathrm{IS} e^{S_0^{(0)}} \right )_l  | \Phi_0 \rangle = 0 \label{eqs01}\\
& \textrm{and} & \nonumber \\
&&  \langle \Phi_v^* | \left \{ \left ( H e^{S_0^{(0)}} \right )_l -E_v^{(0)})  \right \} S_v^{(1)}  + \left ( H e^{S_0^{(0)}} S_0^{(1)} \right )_l \nonumber \\
&&   \times  \left \{ 1+ S_v^{(0)} \right \} + \left ( H_\mathrm{IS} e^{S_0^{(0)}} \right )_l \left \{ 1+ S_v^{(0)} \right \} \nonumber \\
&& + E_v^{(1)} S_v^{(0)}  | \Phi_v \rangle = 0 .  \ \ \ \label{eqv1} \ \ 
\label{eqamp1}
\end{eqnarray}
In the above equation, the expression for an IS factor is given by
\begin{eqnarray}
 E_v^{(1)} &=& \langle \Phi_v | \left ( H e^{S_0^{(0)}} \right )_l  S_v^{(1)}  + \left ( H e^{S_0^{(0)}} S_0^{(1)} \right )_l  \left \{ 1+ S_v^{(0)} \right \}   \nonumber \\
&& + ( H_{IS} e^{S_0^{(0)}})_l \left \{ 1 + S_v^{(0)} \right \} | \Phi_v \rangle . 
\label{eqeng1}
\end{eqnarray}
It should be noted that Eqs.~(\ref{eqs01}),~(\ref{eqamp1}) and~(\ref{eqeng1}) are the first-order approximations of Eqs.~(\ref{eqs0}),~(\ref{eqamp}) and (\ref{eqeng}), respectively. Therefore, the amplitudes of the perturbed operators of the AR-RCC method follow similar expressions corresponding to their respective unperturbed RCC operators.

In this work, we considered correlations from electrons among the 20$s$, 20$p$, 19$d$, 18$f$, 16$g$, 14$h$, and 12$i$ orbitals. Since considering triple excitations among all these orbitals was not feasible with the available computational resources, we have allowed triple excitations up to 15$s$, 15$p$, 15$d$, 11$f$ and 10$g$ orbitals along with the correlations from the RCCSD method. It should be noted that we have not counted spin multiplicity of the orbitals here.

To demonstrate contributions to the IS factors at different levels of approximations in the atomic Hamiltonian, we give results using the DC Hamiltonian and corrections due to the Breit (given as $+$Breit) and QED (given as $+$QED) effects at the RCCSD method. Differences of the RCCSD values from the larger basis set are given as `$+$Basis' contribution. Estimated contributions from the triple excitations are listed as $+$T. We also present calculated energies at the second perturbation theory (MP2 method) using the 20$s$, 20$p$, 19$d$, 18$f$ and 16$g$ orbitals to demonstrate importance of considering an all-order method like RCC theory for accurate calculations of properties in Ag. Most of the uncertainty in our calculated energies and IS factors would stem from the frozen orbitals in the estimations of the triples contributions. High-lying orbitals that are not included in the RCCSD calculations can also contribute to some extent to the IS factors. We have accounted for possible uncertainties from these contributions after estimating them in the MP2 method.

\subsection{Results and Discussion}

\subsubsection{Energies}

In Table~\ref{tab:E} we give results for the calculated EAs.
They are given first at the DHF approximation, which already captures the gross level structure.
When taking into account electron correlations through either MP2 or RCCSD, the EAs of the $n=5$ manifold increase by $\mathcal{O}$($15\%$). The effect is stronger than in the isoelectronic Cd~II, where it is $8-9\%$, indicating that electron correlations are more important in Ag~I.
For the $n=6$ manifold, the increase is half the size, hinting that correlations in more weakly bound single-valence states play smaller roles, as expected.
Introducing correlation effects through triple excitations increases the EAs of the $n=5$ manifold by $0.7-1.5\%$, twice as much as in Cd~II, and three times that for the $n=6$ manifold.
The uncertainty tied to missing quadruple electron excitation contributions to EAs is expected to be small.
The basis set extrapolation increases the EAs for all levels as well.

The Breit and QED contributions are found to be small but not negligible. 
We find that for states with $ns$ valence orbitals, the Breit and QED corrections are comparable in magnitude while for states with the $np$ subshells, the former is much larger, as was pointed out in~\cite{2018-QED}.
Our approximate QED correction to the $5s$$~^2S_{1/2}$ EA is found to be smaller than most other literature values, as compiled in Table~VIII of~\cite{2018-QED}. For this reason, we ascribe a $100\%$ uncertainty to it and to the corresponding corrections to the IS operators.
Although these uncertainties are negligible with respect to other contributions, they point that moving to heavier or multiply-charged systems without losing accuracy would necessitate a refinement of the QED treatment.

The total EAs are $0.2-0.3\%$ away from experiment, 
within two standard deviations from our estimated uncertainty.
They are closer to the experimental values than the closest results from the literature~\cite{2006-theory,2022-The}, which do not quote uncertainties, by up to a factor of $8$. 
Due to the stronger electron correlations, the total uncertainty is a factor $2-3$ higher than in our prior work on Cd~II.
All in all, we undershoot the experimental energies, which indicates a more complete treatment of electron correlations is necessary for accurate estimations of the results. For Cd~II, we overshot the experimental energies, indicating possibly underestimated many-body QED effects.

The EEs are calculated from the differences of the other level EAs to that of the $5s~^2S_{1/2}$ ground level at different approximations. 
Similarly to the EAs, the EEs are $0.2\%$ away from experimental measurements, within our uncertainty estimation. The EEs of the $n=5$ doublet are less accurate than in the prior works~\cite{2022-The,2021-Flam}, while those of the higher states are more so~\cite{2022-The,2006-theory}.
It is interesting to note that contributions from the triple excitations are similar for each state in the manifold. This means that the uncertainty is dominated by basis extrapolation for the first two EEs.

\subsubsection{IS factors}

\begin{table*}[t]
\caption{
Calculated isotope shift factors $F$, $K^\mathrm{SMS}$ and $K^\mathrm{NMS}$ for selected levels in Ag I. For each of the calculated values, we first list factors relative to the ground state of the Ag~II ion, followed by factors for optical transitions from the $5s~^2S_{1/2}$ ground state of Ag I. 
$F$ and $K^\mathrm{SMS}$ are compared with semi-empirical values from the literature, while $K^\mathrm{NMS}$ is compared with the values returned from the scaling law (S.L.),
as discussed in the main text. Calculations are performed with $R=4.55\,$fm, the effect of repeating them with different radii is shown in Table~\ref{tab:EEISA}.
}
\begin{ruledtabular}
\begin{tabular}{l rrr rr r r r}
State  & \multicolumn{1}{c}{DHF}  & \multicolumn{1}{c}{AR-RCCSD} & \multicolumn{1}{c}{+T} & $+$Basis & $+$Breit & $+$QED & Total & \\
\hline \\ 
\multicolumn{1}{l}{\underline{$F$ MHz/fm$^2$}}&&&&&&&& \\
$5s$$~^2S_{1/2}$ & $-2527$& $-3852$& $129(32)$ & $-30(15)$ & $11$ & $19(19)$ & $-3723(40)$  \\
$5p~^2P_{1/2}$ & $-17$  & $-163$ & $-37(9)$  & 1(1)      & 1    & 1(1)     &  $-197(9)$  \\
$5p~^2P_{3/2}$ & $-0.$  & $-126$ & $-42(11)$ & 2(1)      & 1    & 1(1)     & $-165(11)$   \\
$6s$$~^2S_{1/2}$ & $-412$ & $-517$ & $~15(4)$  & $-1(1)$   & 1    & 2(1)     & $-499(5)$  \\
$6p~^2P_{1/2}$ & $-5$   & $-32$  & $-11(3)$  & 1(0.)     & 0.   & 0.(0.)   & $-42(3)$   \\
$6p~^2P_{3/2}$ & $-0.$  & $-25$  & $-15(4)$  & 1(0.)     & 0.   & 0.(0.)   & $-39(4)$ \vspace{1 mm}\\

$5s$$~^2S_{1/2}$$-5p~^2P_{1/2}$ & $-2510$ & $-3689$ & $166(42)$ & $-31(15)$ & $10$ & $19(19)$ & $-3525(48)$  \\
$5s$$~^2S_{1/2}$$-5p~^2P_{3/2}$ & $-2527$ & $-3726$ & $171(43)$ & $-31(16)$ & $10$ & $18(18)$ & $-3557(49)$ \\ 
~~~~ Ref.~\cite{1989-GSI} & \multicolumn{1}{c}{$-2625$,~$-3146$ }&&&&&& $-4265(341)$\\
~~~~ Ref.~\cite{2014-LISOL,2021-Ag} &&&&&&& $-4300(300)$ \\
$5s$$~^2S_{1/2}$$-6s$$~^2S_{1/2}$           & $-2115$ & $-3336$ & $114(29)$ & $-28(14)$ & $10$ & $17(17)$ & $-3223(36)$ \\
$5s$$~^2S_{1/2}$$-6p~^2P_{1/2}$ & $-2522$ & $-3820$ & $141(35)$ & $-30(15)$ & $11$ & $19(19)$ & $-3680(43)$ \\
$5s$$~^2S_{1/2}$$-6p~^2P_{3/2}$ & $-2527$ & $-3827$ & $144(36)$ & $-30(15)$ & $11$ & $19(19)$ & $-3683(43)$ \\
\hline \\
\multicolumn{1}{l}{\underline{$K^\mathrm{SMS}$ GHz~u}} &&&&&&&\\
$5s$$~^2S_{1/2}$ & $-1611$ & $1346$ & $115(29)$ & $-4(2)$ & $5$  & $1(1)$    & $1463(29)$ & \\
$5p~^2P_{1/2}$ & $-553$  & $342$  & $68(17)$  & $-3(2)$ & $-1$ & $-0.(0.)$ & $405(17)$  & \\
$5p~^2P_{3/2}$ & $-464$  & $370$  & $64(16)$  & $-3(1)$ & $-0.$ & $-0.(0.)$ & $432(16)$ & \\
$6s$$~^2S_{1/2}$ & $-253$  & $176$  & $28(7)$   & $-2(1)$ & $1$   & $0.(0.)$  & $202(7)$ &   \\
$6p~^2P_{1/2}$ & $-150$  & $65$   & $23(6)$   & $-1(1)$ & $0.$ & $-0.(0.)$  & $88(6)$  & \\
$6p~^2P_{3/2}$ & $-128$  & $75$   & $25(6)$ & $-1(0.)$  & $0.$  &  $-0.(0.)$ & $99(6)$ \vspace{1 mm}\\

$5s$$~^2S_{1/2}$$-5p~^2P_{1/2}$ & $-1058$ & $1005$ & $47(12)$ & $-1(0.)$& $6$ & $1(1)$   & $1058(12)$  \\
$5s$$~^2S_{1/2}$$-5p~^2P_{3/2}$ & $-1146$ & $976$  & $51(13)$ & $-1(1)$ & $5$ & $1(1)$   & $1031(13)$ \\
~~~~ Ref.~\cite{2014-LISOL} &&&&&&& $150(450)$ \\
$5s$$~^2S_{1/2}$$-6s$$~^2S_{1/2}$ & $-1358$ & $1171$ & $87(22)$ & $-2(1)$ & $4$ & $0.(0.)$ & $1260(22)$ \\
$5s$$~^2S_{1/2}$$-6p~^2P_{1/2}$ & $-1461$ & $1281$ & $92(23)$ & $-3(1)$ & $5$ & $1(1)$   & $1375(23)$ \\
$5s$$~^2S_{1/2}$$-6p~^2P_{3/2}$ & $-1482$ & $1271$ & $90(23)$    & $-3(2)$ & $5$ & $1(1)$   & $1363(23)$ \\
\hline \\ 
\multicolumn{1}{l}{\underline{$K^\mathrm{NMS}$ GHz~u}} &&&&&&&& S.L.\\
$5s$$~^2S_{1/2}$ & 3393 & 808 & $36(9)$  & $11(5)$& $-1$  & $-1(1)$    & $852(11)$ &  1005 \\
$5p ~ ^2P_{1/2}$ & 1200 & 367 & $51(13)$ & $5(3)$ & $-1$  & $0.(0.)$   & $422(13)$ & 519 \\
$5p ~ ^2P_{3/2}$ & 1111 & 351 & $49(12)$ & $5(3)$ & $-1$  & $-0.(0.)$  & $405(13)$ & 504 \\
$6s$$~^2S_{1/2}$ & 667  & 274 & $2(1)$   & $1(1)$ & $-0.$ & $-0.(0.)$  & $277(1)$  &  305  \\
$6p ~ ^2P_{1/2}$ & 396  & 181 & $5(1)$   & $1(1)$ & $-0.$ & $0.(0.)$   & $187(1)$  & 211  \\
$6p ~ ^2P_{3/2}$ & 376  & 176 & $5(1)$   & $1(1)$ & $-0.$ & $-0.(0.)$  & $182(1)$  &  207\vspace{1 mm}\\

$5s$$~^2S_{1/2}$$-5p~^2P_{1/2}$ & $2193$ & $441$ & $-14(4)$ & $5(3)$ & $-0.$& $-1(1)$ & $431(5)$  & 486\\
$5s$$~^2S_{1/2}$$-5p~^2P_{3/2}$ & $2282$ & $457$ & $-13(3)$ & $5(3)$ & $-1$ & $-1(1)$ & $448(4)$  & 501\\
$5s$$~^2S_{1/2}$$-6s$$~^2S_{1/2}$ & $2727$ & $534$ & $34(9)$  & $9(5)$ & $-1$ & $-1(1)$ & $575(10)$ & 700\\
$5s$$~^2S_{1/2}$$-6p~^2P_{1/2}$ & $2998$ & $627$ & $31(8)$  & $9(5)$ & $-1$ & $-1(1)$ & $665(9)$  & 764\\
$5s$$~^2S_{1/2}$$-6p~^2P_{3/2}$ & $3017$ & $632$ & $32(8)$  & $9(5)$ & $-1$ & $-1(1)$ & $671(9)$  & 798 
\end{tabular}
\end{ruledtabular}
\label{tab:factors}
\end{table*}

In Table~\ref{tab:factors}, we give the IS factors evaluated at different levels of approximation. 
We first discuss the FS factors from our calculations.
At the DHF level, $F(5s~^2S_{1/2})$ and $F(6s~^2S_{1/2})$ are large stemming from strong overlap with the nucleus, $F(5p~^2P_{1/2})$ and $F(6p~^2P_{1/2})$ are small but non-negligible, and $F(5p~^2P_{3/2})$ and $F(6p~^2P_{3/2})$ are negligible.
Our result for $F_{328}\equiv$ $F(5p~^2P_{3/2})-F(5s~^2S_{1/2})$ is close to a the Hartree-Fock calculation in~\cite{1989-GSI} but differs from their Dirac-Fock calculation.
Introducing electron correlations at the AR-RCCSD approximation increases $F(5s~^2S_{1/2})$ by $50\%$, and $F(6s~^2S_{1/2})$ by $25\%$. This trend is similar to that seen with the EAs. Correlations also increase $F$ of the $nP$ levels, thus reducing magnitudes for the fine-structure intervals.
Triple excitation contributions to $F(5s~^2S_{1/2})$ reduce its magnitude by $3\%$, twice as much and of opposite sign as in Cd~II. In Zn~II, triple contributions to $F(4s~^2S_{1/2})$ are of the same sign and five times smaller then in Ag~I~\cite{2023-Zn}.
These observations demonstrate the non-trivial nature of correlation effects from the high-level excitations, and motivate extending these calculations to the homologous levels in Cu~I in order to gain further insight.
As in Cd~II and Zn~II, the contribution of approximate QED corrections is much larger than that of the Breit interaction. Nevertheless, it is still small compared with our uncertainty estimation.

To our knowledge, ours is the first \textit{ab initio} calculation of the FS factor in Ag~I levels.
Nevertheless, it was semi-empirically extracted from the $5s$$~^2S_{1/2}$ hyperfine structure, yielding $F_{328,\mathrm{SE}}=-4265(341)$~MHz/fm$^2$, with `SE' shorthand for semi-empirical, given in~\cite{1989-GSI}. A similar estimation was also used in recent works~\cite{2014-LISOL,2021-Ag}.
$F_{328,\mathrm{SE}}$ is larger than our recommended value by two of its standard deviations.
A quarter of the difference is directly related to the missing contribution from the $5p$ level, and another quarter from their estimation of the higher-moment contribution.
These observations support other studies (see~\cite{2003-Julian,2006-Magda,2019-Ne}) which suggest that semi-empirically estimated values of $F$ are about 20\% too large. Whilst this effect was already reported by Torbohm \textit{et al} in 1985~\cite{1985-Torbohm}, semi-empirical values for $F$ are still often assigned a much smaller error in the literature.

\begin{table*}[t]
\caption{
Determining the dependence of energies and isotope shift factors on the assumed root mean-squared nuclear charge radius, $R$, in fm.
Calculated values of electron attachments and isotope shift factors in different states with different values for the nuclear charge radius. The Dirac-Coulomb Hamiltonian at the Analytical-response, relativistic coupled-cluster up to double excitations [(AR-)RCCSD] approximation is used. The R=4.55 column corresponds to the AR-CCSD column in Table~\ref{tab:factors}.
}
\begin{ruledtabular}
\begin{tabular}{l c c c  c  c}
     &  $R=4.34$ &  $R=4.54$  & $R=4.55$ & $R=4.56$ & $R=4.84$ \\
\hline \\ 
\multicolumn{1}{l}{\underline{EA values in cm$^{-1}$}} \\
$5s$$~^2S_{1/2}$ & 60407.91 & 60407.69  & 60407.67 & 60407.66 & 60407.33 \\
$5p~^2P_{1/2}$ & 31007.44  & 31007.43 & 31007.43 & 31007.43 & 31007.42 \\
$5p~^2P_{3/2}$ & 30088.87 & 30088.86  &  30088.86 & 30088.86 & 30088.85 \\ 
$6s$$~^2S_{1/2}$ & 18454.78 & 18454.75 & 18454.75 &  18454.74 & 18454.70 \\
$6p~^2P_{1/2}$ & 12680.42 & 12680.41 &  12680.41 & 12680.41 & 12680.41 \\
$6p~^2P_{3/2}$ & 12467.25 & 12467.25 & 12467.25 & 12467.25 & 12467.25 \\
\hline \\
\multicolumn{1}{l}{\underline{$F$ MHz/fm$^2$}} \\
$5s$$~^2S_{1/2}$ & $-3866.85$ & $-3853.40$  & $-3852.20$ &  $-3850.63$  & $-3830.17$ \\
$5p~^2P_{1/2}$ & $-163.99$ & $-163.38$  &  $-163.39$ & $-163.26$  & $-162.40$  \\
$5p~^2P_{3/2}$ & $-126.66$ & $-126.22$  &  $-126.25$ & $-126.14$ & $-125.47$ \\ 
$6s$$~^2S_{1/2}$ & $-518.63$   & $-516.82$  &  $-516.66$  &  $-516.45$ & $-513.71$ \\
$6p~^2P_{1/2}$ & $-32.03$ & $-31.92$ &  $-31.92$ &  $-31.89$ & $-31.73$ \\
$6p~^2P_{3/2}$ & $-25.33$ & $-25.24$ &  $-25.24$  & $-25.22$ & $-25.09$\\
\hline \\
\multicolumn{1}{l}{\underline{$K^\mathrm{NMS}$ GHz~u}} \\
$5s$$~^2S_{1/2}$  & 807.82 & 807.79 & 807.88 & 807.77 & 807.73 \\
$5p~^2P_{1/2}$ & 366.64 & 366.63 & 366.68 & 366.63 & 366.67  \\
$5p~^2P_{3/2}$ & 350.68  & 350.67 & 350.71 & 350.67 &  350.71  \\
$6s$$~^2S_{1/2}$ &  273.87  & 273.86  & 273.87 & 273.86 & 273.85  \\
$6p~^2P_{1/2}$ & 180.86 & 180.86 & 180.87 & 180.86 & 180.87 \\
$6p~^2P_{3/2}$ & 175.74  & 175.74 &  175.75 & 175.74  &  175.75 \\
\hline \\ 
\multicolumn{1}{l}{\underline{$K^\mathrm{SMS}$ GHz~u}} \\
$5s$$~^2S_{1/2}$ & 1346.77 & 1346.49 & 1346.33 & 1346.49  & 1346.41 \\
$5p~^2P_{1/2}$ & 342.32 & 341.77 & 341.68 & 341.72 & 341.79 \\
$5p~^2P_{3/2}$ & 370.81 & 370.27  & 370.16  & 370.23 & 370.22 \\
$6s$$~^2S_{1/2}$ & 175.67 & 175.61  &  175.59 & 175.60 & 175.60 \\
$6p~^2P_{1/2}$ & 65.30 & 65.19  &  65.17 & 65.18 & 65.18 \\
$6p~^2P_{3/2}$ & 75.45 & 75.34 & 75.32 & 75.33 & 75.33 \\
\end{tabular}
\end{ruledtabular}
\label{tab:EEISA}
\end{table*}

The SMS factor $K^\mathrm{SMS}$, tied with a two-body operator, is entirely affected by electron correlations. So much so that its calculated value is meaningless at the mean field approximation.
At the AR-RCCSD approximation, $K^\mathrm{SMS}(5s~^2S_{1/2})$ is of similar magnitude in Ag~I and Cd~II. However, $K^\mathrm{SMS}(5p)$ for both levels of the doublet is $2.5$ times larger in Ag. Triple excitation contributions are twice larger for $K^\mathrm{SMS}(5s~^2S_{1/2})$ and an order of magnitude larger for both $K^\mathrm{SMS}(5p~^2P_{1/2})$ and $K^\mathrm{SMS}(5p~^2P_{3/2})$
Surprisingly, $K^\mathrm{SMS}$ converges fast with increasing basis size. This is fortunate as the calculation of $K^\mathrm{SMS}$ at the AR-CCSDT approximation already requires several months of computation time on a medium-sized high-performance cluster.
As in Cd~II, Breit and QED corrections to $K^\mathrm{SMS}$ are completely negligible.
Thus the total uncertainty is dominated by our estimation of contributions from the missing quadruple excitations. 

To our knowledge, there aren't any available \textit{ab initio} calculations of $K^\mathrm{SMS}$ to be compared with. Nevertheless, we can test the heuristic used in~\cite{2014-LISOL}, that $K^\mathrm{SMS}_\mathrm{SE}\approx0.3(9)K^\mathrm{NMS}_\mathrm{S.L.}=150(450)\,$GHz~u, where the NMS factor was estimated through the scaling-law (S.L.) $K^\mathrm{NMS}_\mathrm{S.L.}=-m_e\Delta E=501\,$GHz~u, with $\Delta E$ the measured energy difference.
Even though a large uncertainty is attached to this semi-empirical calculation, our result lies two standard deviations away, putting some doubts on the reliability of using this semi-empirical method.

Although $K^\mathrm{NMS}$ is estimated from a one-body operator, it is highly affected by electron correlations, with the AR-CCSD value only a third of the DHF value. 
Triple excitations contribute around $10\%$ to the $n=5$ manifold states, as seen in Cd~II. 
For the $n=6$ manifold, the effect of correlations is smaller; half and a few percent in the AR-RCCSD method and after including contributions from the triple excitations, respectively.
Breit contributions to $K^\mathrm{NMS}$ are found to be small but not negligible, while QED contributions are negligible.
A difference of $10$ to $20\%$ between the calculated $K^\mathrm{NMS}$ and the one semi-empirically estimated via scaling-law is observed for all levels. It is much larger than the difference seen in Cd~II, indicating that its origin is from electron correlations, stronger in Ag~I, than from the relativistic effects, stronger in Cd~II. 
A discussion on this phenomenon can be found in Ref.~\cite{1987-Palmer}.

The calculations are repeated for different values of $R$ in a range spanning the root-mean-squared radii of $^{96}$Ag$-^{121}$Ag. The results are given in Table~\ref{tab:EEISA} and show that the dependency of the calculated IS factors on $R$ is negligible compared with our reported uncertainties. Nevertheless, we take them into account when extracting $\delta r^2$ away from stability.

\clearpage
\newpage

\section{Measurements on the $\mathbf{5s~^2S_{1/2} \rightarrow 5p~^2P_{3/2}}$ line at 328 nm}

\begin{figure*}
    \centering
    \includegraphics[trim=0 2 1 6,clip,width=0.84\textwidth]{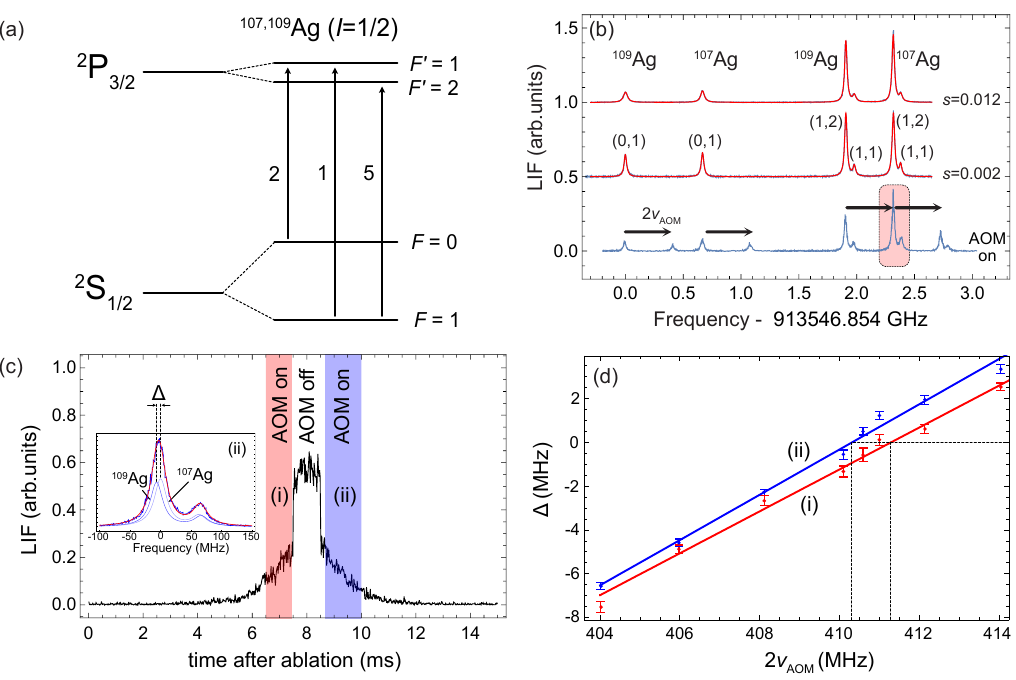}
    \caption{Laser induced fluorescence spectroscopy of a buffer gas cooled Ag atomic beam. (a) Level scheme for the $^2S_{1/2}\rightarrow{} ^2P_{3/2}$ transition in $^{107,109}$Ag. The relative intensities of the transitions are indicated.
    (b) Example spectra taken using a single frequency probe beam with $s=0.012$ (upper), a single frequency probe beam with $s=0.002$ (middle), and a two-frequency probe beam whose frequencies are separated by $2\nu_{\mathrm{AOM}} \approx 410$~MHz (lower). Resonance lines are labelled by isotope and total angular momentum numbers ($F,F'$) for the transition. The red shaded box indicates the region where the (1,2) transitions of $^{107,109}$Ag are almost overlapping.
    (c) Example time-of-flight trace of the fluorescence when using the double-pass acousto-optic modulator (AOM) as discussed in the text, illustrating the observation windows used in the analysis. A sudden change in fluorescence signal occurs when the AOM is switched on or off. The inset shows a spectrum using observation window (ii), and the fitted frequency offset $\Delta$ between the (1,2) lines of the Ag isotopes. The contributions of the individual isotopes are shown in transparent blue, with the solid red line their sum which is the combined fit function.
    (d) A plot of the fitted value of $\Delta$ versus $2\nu_{\mathrm{AOM}}$ for observation windows (i) and (ii), used to extract the isotope shift of the (1,2) lines.}
    \label{fig:expData}
\end{figure*}

To improve the available measurements for the naturally occurring isotopes of Ag, we perform continuous-wave laser induced fluorescence spectroscopy with a buffer gas cooled atomic beam. The spectrometer used for these measurements has been described previously~\cite{2022-Doppelbauer,2023-Cd,2023-Zn2}. Briefly, Ag atoms of natural isotopic abundance ($52\%$ $^{107}$Ag and $48\%$ $^{109}$Ag) are produced by laser ablation inside a cryogenically-cooled copper cell, thermalise by colliding with He buffer gas at a temperature of 3~K, and exit the cell as a slow, pulsed atomic beam. At a distance of 70~cm from the exit of the buffer gas cell, the atoms interact with a low intensity probe laser beam which excites the $5s~^2S_{1/2}\rightarrow 5p~^2P_{3/2}$ transition near 328~nm. The laser light is produced by frequency doubling a narrow-linewidth ring dye laser (Sirah Matisse 2DX) at 656~nm. The frequency of the 656~nm light is recorded using a commercial wavemeter (High Finesse WS8-10, calibrated with a temperature stabilised HeNe laser) which provides an absolute accuracy of 20~MHz at~328 nm. Laser induced fluorescence is collected using a photomultiplier tube whose photocurrent is delivered to a transimpedance amplifier to generate a time-of-flight trace. Based on the range of arrival times at the detector, we estimate that the range of velocities in the beam covers 90 to 130~m/s (full-width at half-maximum). The transverse velocity width of the atomic beam is restricted using a 2~mm square aperture mounted immediately in front of the detector. Orthogonality with the atomic beam direction is ensured using a set of alignment irises mounted on the detection vacuum chamber; we additionally verify and limit residual Doppler shifts due to misalignment by comparing spectra by arrival time at the detector, as discussed later. 

Figure \ref{fig:expData}a shows the relevant energy levels for $^{107,109}$Ag, both of which have a nuclear spin of $I=1/2$. Levels within the ground (excited) levels are labelled by their hyperfine angular momentum quantum number $F$ ($F'$), and the optical transitions are labelled by their respective line intensities.  In the experiment, the hyperfine structure in the $^{2}S_{1/2}$ and $^2P_{3/2}$ states is considered. The large splitting ($\sim 1.8$~GHz) in the $5s$ ground state means that atoms excited on the $F=0\rightarrow F'=1$ and $F=1 \rightarrow F'=1$ hyperfine transitions are optically pumped to the $F=1$ ($F=0$) levels after an average of 2 (3) photon scattering events, respectively. The process of optical pumping tends to reduce the amplitudes and increase the width of these lines as compared to the closed $F=1\rightarrow F'=2$ transition, and necessitates minimizing the number of photon scattering events per atom to obtain the best spectra.
Secondly, the hyperfine splitting in the $5p~^2P_{3/2}$ state is roughly three times the natural linewidth of the transition, which modifies the fluorescence lineshape by the interference between photon scattering pathways~\cite{2013-Brown}. To minimize this effect, we use laser light linearly polarized at an angle $\theta_{\mathrm{m}} = \cos^{-1}(1/\sqrt{3})$ to the detector direction. This is the so-called ``magic angle" at which the anisotropic part of the fluorescence emission is zero, interference between scattering paths disappears, and a symmetric lineshape is recovered~\cite{2013-Brown}. The relatively large solid angle of our collection optics of nearly $\pi/4$ steradians further suppresses the interference effect by roughly a factor 2. 

Figure~\ref{fig:expData}b shows three spectra obtained with our spectrometer. The upper two spectra are taken with a single-frequency probe laser, at two different probe laser intensities $I_0$. We label the spectra by the two-level saturation parameter $s= I_0/I_{sat}$, where $I_{\mathrm{sat}}= \frac{\pi h c \Gamma}{3\lambda^3} = 82.5$~mW/cm$^2$ is the two-level saturation intensity for the transition. Solid red lines show fits using a set of Lorentzian functions, where the full-width at half maximum $\Gamma/(2\pi)$ is allowed to vary between resonances in order to account for optical pumping. For the spectrum at lower intensity, the relative line intensities agree well with the relative line intensities given in panel~(a), indicating that optical pumping has been largely avoided. For the higher intensity spectrum, we use the relative peak heights to estimate that when the laser is tuned to the $(F,F') = (1,2)$ resonance, the atoms scatter on average six photons. This means that the effect of photon recoil shifts and broadens the resonance lines by at most 0.1~MHz. We note that the number of photon scattering events derived from the line intensities is a factor two (less) than derived using a simple two-level rate equation model and the estimated intensity of the probe light. The Lorentzian linewidth of the $(1,2)$ resonances, being unaffected by optical pumping, is $\Gamma/(2\pi)=25.4(1.0)$~MHz. Fitting with a Voigt lineshape resulted in a slightly reduced Lorentzian linewidth, $\Gamma/(2\pi)=24.4(6)$~MHz, with a Gaussian linewidth of below 4~MHz (full width at half maximum), corresponding to a transverse velocity spread of below 3.1~m/s. The radiative lifetime extracted from the Voigt fits, $\tau = 1/\Gamma$, is a few percent below that reported by Carlsson et al. \cite{1990-Carlsson}, which we attribute to the effect of a residual magnetic field in our detector, measured to be $\sim0.3$~G.

The absolute frequencies of the line centres from six spectra, taken over two days and using a range of probe laser intensities, varied by less than 2~MHz (standard deviation), and intervals between resonance lines varied by less than 1.5~MHz (standard deviation). The absolute frequency uncertainty is dominated by the 10~MHz uncertainty of the wavemeter at 656~nm; in a previous experiment using the same spectrometer we found agreement with the precisely measured $^1S_0 \rightarrow{} ^1P_1$ 399~nm line in Yb at the 10~MHz level~\cite{2022-Doppelbauer}. Doppler shifts due to misalignment of the probe laser and atomic beams is at most 2~MHz, which we estimate by plotting the fitted line centres versus forward velocity in the atomic beam, and extrapolating to zero velocity. 
We extract the magnetic dipole hyperfine interaction constants $A_{107}$ and $A_{109}$ in each state for the two Ag isotopes and present these in Table \ref{tab:ExpData}. The ground state splittings agree with high accuracy microwave measurements~\cite{1953-Wessel, 1967-Dahmen} to within 2.5~MHz ($^{109}$Ag) and 0.4~MHz ($^{107}$Ag), and for the excited state we agree with quantum beat spectroscopy measurements of Carlsson et al.~\cite{1990-Carlsson} to better than 0.4~MHz. This suggests that the isotope shift between $^{107,109}$Ag measured from the same spectra may have an uncertainty of 2-3 MHz due to the uncertainty of the wavemeter frequency measurements.

To improve the accuracy of our isotope shift measurements, we introduced a single acousto-optical modulator (AOM) into the optical setup, operated in double-pass configuration using a cat-eye lens. The AOM is driven at a radio frequency $\nu_{\mathrm{AOM}}$, which can be varied via a voltage controlled oscillator, and is monitored using a radio frequency counter. We deliberately allow both the zero-th order (i.e. unshifted) and twice-shifted beams to be present, such that the probe laser light is now composed of two frequency components separated by $2\nu_{\mathrm{AOM}}$, and whose spatial overlap and pointing varies by less than 0.4~mrad as $\nu_{\mathrm{AOM}}$ is varied. An example spectrum is shown in the lower trace of figure \ref{fig:expData}b, where arrows show the displacement in the frequency axis introduced by the AOM. We choose $\nu_{\mathrm{AOM}}$ to be around 200~MHz such that the frequency-shifted component of the probe beam excites the $^{107}$Ag, $F=1-F'=2$ transition, whilst the other component simultaneously excites this transition in $^{109}$Ag. This enables detecting both isotopes whilst only scanning a small ($\sim$200~MHz) range with the laser, indicated by the red shaded box in the spectrum. Moreover, we can rapidly switch between single and dual-frequency probe light as atoms fly through the detector via the RF driving power to the AOM. 

Figure \ref{fig:expData}c shows a time of flight fluorescence trace using the AOM method, and with the laser tuned to the $(1,2)$ $^{107}$Ag resonance. The RF power to the AOM is set to give equal optical power in the two frequency components, and is switched off for roughly 200~$\mu$s as the atoms fly through the detector. We use the fluorescence within this observation window for a ``reference spectrum" in order to fix the $^{107}$Ag isotope position in each measurement. This largely removes contributions to the IS from any slow drift in the wavemeter frequency axis over timescales longer than a minute. The shaded regions in the figure labelled (i) and (ii) are observation windows for which both frequency components are present in the probe light. We fit spectra derived from these observation windows to a model containing the resonances of both isotopes, where the hyperfine splittings are fixed to those given in Table~\ref{tab:ExpData}, the $^{107}$Ag (1,2) resonance is fixed to that measured in the reference spectrum, and the natural abundance ratio determines the total contribution from each isotope to the spectrum. This leaves only the offset frequency $\Delta$ between the (1,2) resonances of $^{107,109}$Ag as a free parameter, beside a common amplitude term and a $y$-axis offset. The inset to Fig.~\ref{fig:expData} displays an example spectrum with $2\nu_{\mathrm{AOM}} = 404.02$~MHz, showing the two isotope components of the signal and the interval $\Delta$ in the underlying fit function. 

\begin{table}[!tbp]
\caption{
Summary of the experimental results for the $5s$$~^{2}S_{1/2}\rightarrow{}$ $5p$$~^{2}P_{3/2}$ transition in $^{107,109}$Ag,
and comparison with literature values.
All values are given in MHz. Absolute frequencies of hyperfine lines for isotope $\alpha$ are labelled as $\nu_{\alpha}(F,F')$ and $\bar\nu_{\alpha}$ denotes the gravity centre for this isotope. 
$\bar{\nu}_\mathrm{nat.}$ is the mean of the gravity centers for the two isotopes, weighted by the natural isotopic abundance, which approximates the observed line-center in a low-resolution measurement with a naturally abundant sample.
$\delta\nu_{109,107}(1,2)=\nu_{107}(1,2)-\nu_{109}(1,2)$. $\delta\bar{\nu}_{109,107}=\bar{\nu}_{107}-\bar{\nu}_{109}$.
} 
\begin{ruledtabular}
\begin{tabular}{l l l }
\multicolumn{1}{l}{} & \multicolumn{1}{l}{This work} & Literature \\
$\nu_{109}(1,2)$ & $913~548~760(20)$& \\
$\nu_{107}(1,2)$ & $913~549~171(20)$& \\
$A_{109}(^2P_{3/2})$ & $-36.9(3)$ & $-36.7(7)$~\cite{1990-Carlsson}  \\
$A_{109}(^2S_{1/2})$ & $-1979.4(1.1)$ & $-1976.932~075(17)~$\cite{1967-Dahmen} \\
$A_{107}(^2P_{3/2})$ & $-31.7(6)$ &  $-31.7(7)$~\cite{1990-Carlsson}  \\
$A_{107}(^2S_{1/2})$ & $-1713.0(8)$ & $-1712.512~111(18)$~\cite{1967-Dahmen} \\
$\bar{\nu}_{109}$ & $913~548~293(20)$ & \\
$\bar{\nu}_{107}$ & $913~548~766(20)$ & \\
$\bar{\nu}_\mathrm{nat.}$ & $913~548~539(20)$ & $913~548~593(60)$~\cite{2001-absolute}\\
$\delta\nu_{109,107}(1,2)$ & $410.9(6)$ & \\
$\delta\bar{\nu}_{109,107}$ & $473.2(7)$ & $467(4)$~\cite{2021-Ag}\\
 & & $476(10)$~\cite{2000-MOT}\\
\end{tabular}
\end{ruledtabular}
\label{tab:ExpData}
\end{table}

We repeat this procedure for different values of $2\nu_{\mathrm{AOM}}$ and plot $\Delta$ versus $2\nu_{\mathrm{AOM}}$ in Fig.~\ref{fig:expData}d. In the ideal case, this should result in a linear relationship with a gradient 1, whose $x$-axis intercept (i.e. where $\Delta =0$) returns the isotope shift between the $(1,2)$ lines of the two Ag isotopes. The linear fits to the data in observation windows (i) and (ii) have slightly different intercepts, which we attribute to a small residual misalignment of the probe light to the atoms, and the atoms in the two observation windows having slightly different forward velocities. The slopes of the two curves are $0.96(2)$ and $1.03(2)$ in regions (i) and (ii) respectively. We take the average of the two intercepts as the $x$-axis intercept, and half their difference as the 67\% confidence interval, and apply small corrections to this value to arrive at our value for the true isotope shift as follows. Firstly, we consider the residual alignment difference between the two frequency components in the probe light, which for a forward velocity of 120~m/s tends to reduce the line separation by 150~kHz. Secondly, the spectral intensity of the probe light near the $^{107}$Ag resonance is about a factor 4 larger (i.e. $s\approx 0.016$) in the reference spectrum, which would shift the $(1,2)$ line centre of $^{107}$Ag in the reference spectrum by +70~kHz. We both correct for these effects and increase the systematic error bar accordingly. Thirdly, the background magnetic field in the detector ($\sim 0.3$~G) leads to Zeeman shift-induced broadening of the hyperfine lines and potentially to a small differential shift. Since the nuclear spins of $^{107,109}$Ag are identical, and the nuclear magnetic moments are within about $15\%$ of one another \cite{2005-STONE}, this effect is negligible. AC Stark shifts of the $^2P_{3/2}$ and $^2S_{1/2}$ states by the excitation light are well below the~kHz level for the intensities used in our measurements and can be neglected. The result, $\delta\nu_{109,107} = \nu_{107}(1,2) - \nu_{109}(1,2)=410.9(6)$MHz, is consistent with that measured using the wide range, single frequency probe data, $410.7(7)$~MHz, though the latter has a few MHz systematic uncertainty associated with the linearity of the wavemeter.
Our procedure is robust to relaxing many assumptions about the underlying lineshape model. For example, fitting the overlapping $(1,1)$ resonances as a single Lorentzian line changes $\delta\nu_{109,107}(1,2)$ by less than 0.1~MHz. Varying the contribution of $^{107,109}$Ag to the overlapping $(1,2)$ lines also results in a value for $\delta\nu_{109,107}(1,2)$ consistent within 0.1~MHz. Such a signal imbalance only significantly changes the slope observed in Fig.~\ref{fig:expData}d.

In deriving an improved value for the isotope shift of the $5s~ ^2S_{1/2} \rightarrow 5p$ $^2P_{3/2}$ gravity centre, we take advantage of the $^2S_{1/2}$ hyperfine-structure measurements of~\cite{1967-Dahmen}, whose stated uncertainty is below 0.1~kHz. We use the values in Table \ref{tab:ExpData} for the hyperfine splitting of the $^{2}P_{3/2}$ states, and use the result of Fig.~\ref{fig:expData}d to fix the isotope shift of the $(1,2)$ lines. This interval contributes most to the uncertainty of the gravity centre isotope shift, and therefore dominates the error bar.

\section{Mean-squared Radius difference}

\subsection{Between stable isotope pairs}\label{Sys}

\begin{table}[!tbp]
\scriptsize
\caption{
Extraction of the mean-squared radius difference, $\delta r^2_{109,107}\equiv r^2_{107}-r^2_{109}$ in fm$^2$, of the stable isotopes of Ag via different optical transitions from the ground state.
$\delta \nu_{109,107} \equiv \nu_{107}-\nu_{109}$ are the center of gravity isotope shifts in MHz, estimated from the indicated data sources, including this work (TW). 
The corresponding $\delta r^2_{109,107}$ in fm$^2$ are extracted employing the factors given in Table~\ref{tab:factors}.
The uncertainties are denoted with subscripts according to: `exp' - experiment, `$K,F$' - uncertainties in our calculated isotope shift factors, `std' - standard deviation from unweighted mean,
`NS' - systematic uncertainty from nuclear shape variation, 
and `NP' - uncertainty in extraction from muonic atoms due to nuclear polarization.} 
\begin{ruledtabular}
\begin{tabular}{l l l l }
\multicolumn{1}{l}{Interval} & \multicolumn{1}{l}{Source} & $\delta \nu_{109,107}$ & \multicolumn{1}{c}{$\delta r^2_{109,107}$}\vspace{2mm}\\
$5s$$~^2S_{1/2}$$-5p~^2P_{1/2}$ & \cite{1968-1937,2004-TP}  &$473(4)$          & $-0.207(1)_\mathrm{exp}(3)_{K,F}$\\
$5s$$~^2S_{1/2}$$-5p~^2P_{3/2}$ & TW      &$473.2(7)$                          & $-0.204(0)_\mathrm{exp}(3)_{K,F}$\\
$5s$$~^2S_{1/2}$$-6s$$~^2S_{1/2}$ & TW,~\cite{1962-Fischer,1968-1937,2004-TP}   &$368(7)$  & $-0.212(2)_\mathrm{exp}(3)_{K,F}$\\
$5s$$~^2S_{1/2}$$-6p~^2P_{3/2}$ & \cite{2001-548,2006-2P} &$414.0(6)$          & $-0.207(0)_\mathrm{exp}(3)_{K,F}$ \vspace{1mm}\\
\multicolumn{2}{l}{\textbf{Final}} & \multicolumn{2}{r}{$-0.207(3)_\mathrm{std}(3)_{K,F}(4)_\mathrm{NS}$} \\
\multicolumn{1}{l}{Muonic Ag} &\cite{2004-FH} & \multicolumn{2}{r}{$-0.198(0)_\mathrm{exp}(4)_\mathrm{NP}(5)_\mathrm{NS}$} \\
\multicolumn{1}{l}{Interpolated}&\cite{1989-GSI} & \multicolumn{2}{c}{$-0.148(31)$}\\
\multicolumn{1}{l}{Compilation}&\cite{2013-AM} & \multicolumn{2}{c}{$-0.148(1)$}
\end{tabular}
\end{ruledtabular}
\label{tab:Ritz}
\end{table}

We now combine our calculations of $F$ and $K$ with optical isotope shifts to estimate the mean-squared charge radius difference of the stable isotope pair, $\delta r^2_{109,107}$.
The calculated relative transition field shift factors, $F(i)-F(5s~^2S_{1/2})$, span a range of $-3223(26)$ to $-3683(43)$~GHz/fm$^2$; an order of magnitude larger than their individual uncertainties (see Table~\ref{tab:factors}).
Thus, a useful consistency check of our $F$ and $K$ calculations, as well as the experimental ISs, is satisfied when applying Eq.~(\ref{eq:IS}) to each optical transition returns similar values of $\delta r^2_{109,107}$.
The relative ISs are calculated with a Ritz-type analysis (see e.g.~\cite{2019-Ne}) of our measurement and those given in~\cite{1962-Fischer, 1968-1937, 2001-548, 2004-TP, 2006-2P}. 
The results are given in Table~\ref{tab:Ritz}.

The standard deviation of the radii extracted from individual transitions is $0.003\,$fm$^2$. It is slightly larger than the $0.002\,$fm$^2$ which we would expect from the uncorrelated uncertainties.
This could indicate possible underestimated uncertainties in the experiment and or our calculation.
To account for it, we conservatively add the standard deviation as another source of systematic uncertainty to all of our recommended values of $\delta r^2$.

Our final recommended value for $\delta r^2_{109,107}$ including the above uncertainties 
is within $1.0\,$ combined standard uncertainties from
$\delta r^2_{109,107}=(R_k^{109}/V_2^{109})^2-(R_k^{107}/V_2^{107})^2$, where the Barret-equivalent moments $R_k$ are measured with muonic atom X-ray spectroscopy~\cite{2004-FH}, and the proportionality factors between the Barrett and second charge moment $V_2\equiv R_k/R$ are estimated with a Fermi distribution, to which we added a relative shape-variation uncertainty. 
It is illuminating to note that in both muonic and electronic silver, it is the (unknown) shape change which dominates the error in the extracted radii.
This motivates to perform an elastic electron scattering experiment to determine the shape change.
The marginal agreement between radii differences extracted all-optically and from muonic atoms is 
expected considering recent work in the medium-mass region~\cite{2016-Cu,2022-Cd,2023-Zn}, pointing to the need to reanalyze the cascade energies with modern tools (e.g.~\cite{2005-Paul,2013-Indel,2016-pion,2017-NAt,2022-Muonic,2022-Nat,2022-NAt2,2024-VP}).
It is also worth noting that in Ref.~\cite{2021-Ag}, a more conservative uncertainty estimate for the muonic data was employed.

\begin{figure}[!b]
\centering
\includegraphics[trim=44 55 68 32,clip,width=0.999\columnwidth]{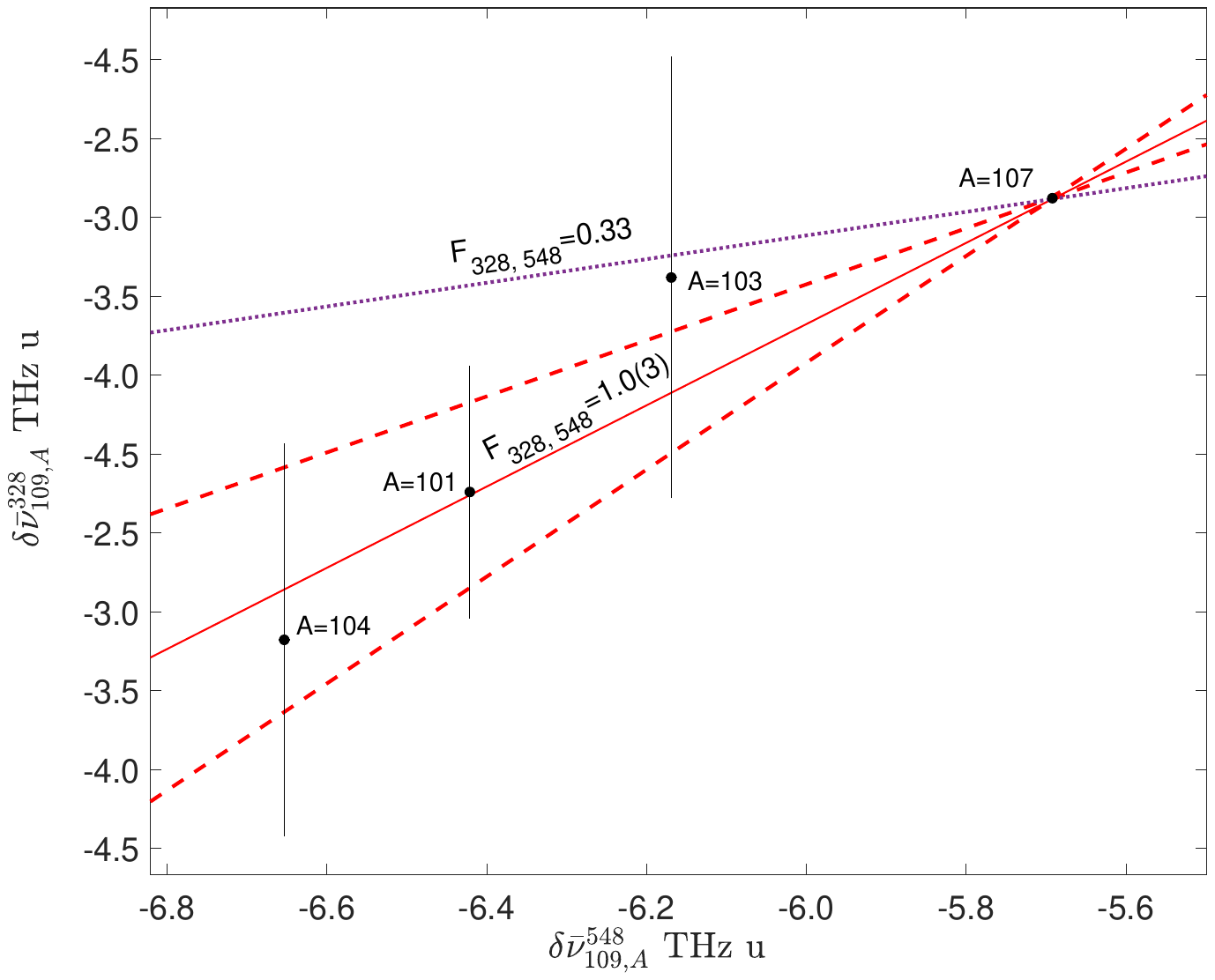}
\caption{Two-dimensional King Plot whose data-points are given in bold in Table~\ref{tab:rad}. 
The vertical axis is the reduced IS in the $5s~^2S_{1/2}-5p~^2P_{3/2}$ transition.
The horizontal axis is the reduced isotope shift in the $4d^95s^2~^2D_{5/2}-6p~^2P_{3/2}$ transition.
The resulting slope (full line) lies two standard deviations (dashed lines) away from that calculated empirically (dotted line) in~\cite{1989-GSI}.}
    \label{fig:Fit}
\end{figure}

We conclude this section by focusing on the IS of the fine-structure of the $5p$ doublet. 
Calculating it from the factors of Table~\ref{tab:factors} and the recommended $\delta r^2_{109,107}$ from Table~\ref{tab:rad} results in $-8.3(4)\,$MHz; in tension with the experimental value of $0\pm4\,$MHz, whose uncertainty is completely decided by a single photoelectric measurement~\cite{1968-1937}.
Extending precision laser spectroscopy to the IS of the $5s$$~^2S_{1/2}$$-5p~^2P_{1/2}$ $338$\,nm line could help to shed light on this issue.
The experimental method detailed in this article could be straightforwardly applied.
 
\subsection{Among isotope and isomer chains}

We interpret the ISs measured in radioisotopes of Ag in terms of $\delta r^2$.
The nuclei are divided into three groups. 
The first consists of those whose ISs were measured only for the $5s~^2S_{1/2}- 5p~^2P_{3/2}$ $328~$nm line (Refs.~\cite{1975-108,2014-LISOL,2021-Ag}). Here, $\delta r^2_{109,A}$ can be estimated directly using our calculated IS factors, including the systematic
uncertainties discussed in the previous subsection.
The results are given in Table~\ref{tab:rad}.
The second group includes the four nuclei whose ISs were measured for both $328\,$nm and $4d^95s^2~^2D_{5/2}-6p~^2P_{3/2}$ $548\,$nm lines (data from this work and Refs.~\cite{2001-548,1989-GSI,2014-LISOL,2021-Ag}, given in bold in Table~\ref{tab:rad}). 
Because the $548\,$nm line involves a $4d^9 5s^2$ configuration, i.e. a 4$d$-hole state, calculating $F_{548}$$\equiv F(6p~^2P_{3/2})-F(4d^95s^2~^2D_{5/2})$ and $K_{548}$$\equiv K(6p~^2P_{3/2})-K(4d^95s^2~^2D_{5/2})$ accurately is beyond the scope of this article. 
We interpret these data by making use of a two-dimensional KP linear equation
\begin{eqnarray}
\delta \bar{\nu}_{A,A'}^{328} &\simeq& 
K_{328,548} + F_{328,548}~\delta \bar{\nu}_{A,A'}^{548},
\label{eq:KP}
\end{eqnarray}
where $\delta \bar{\nu}_{A,A'}^{i}\equiv\delta \nu_{A,A'}^{i}/\mu_{A,A'}$, $F_{328,548}\equiv F_{328}/F_{548}$ and $K_{328,548}\equiv K_{328}-F_{328,548}K_{548}$. 
The higher order corrections to Eq.~\ref{eq:IS} affect the validity of Eq.~\ref{eq:KP}. The largest effect stems from the variation in nuclear shape, which is of order of $2\%$ percent of the field shifts of the two transitions. Thus the maximal change in the slope $F_{328,548}$ is $4\%$, negligible compared with the statistical uncertainty in it, which is $30\%$.
A Monte-Carlo linear regression, shown in Fig.~\ref{fig:Fit}, returns a slope $F_{328,548}$$=1.0(3)$, and intercept $K_{328,548}$$=3.4(1.9)$~THz~u.
It also results in posterior ISs for the cooling line, which we denote $\delta\nu_{109,A}^{328, \mathrm{KP}}$, given in Table~\ref{tab:rad}, along with their corresponding radii.
The third group consists of nuclei for which measurements exist only for the $548\,$nm line. We use the joint distribution of the slope and intercept to project their ISs from the $548\,$nm line to the $328\,$nm line. The results are also given in Table~\ref{tab:rad}, with the corresponding radii.

The fit results can be used to check for the reasons for inconsistencies found in the literature.
The fitted slope deviates by two of its standard deviations from $F^\mathrm{SE}_{548,328}$ $F^\mathrm{SE}_{328,548}$$=0.33(3)$, estimated from the semi-empirical FS factors given in~\cite{1989-GSI}.
Combining our calculated factors for the cooling line with the fitted slope and intercept, we find $F_{548}=-3^{+1}_{-2}~$GHz/fm$^2$ and
$K_{548}=-2^{+2}_{-1}~$THz~u.
Although it is roughly estimated, our FS factor  
is four combined standard errors away from the semi-empirical estimation $F_{548}^\mathrm{SE}=-12(1)~$GHz/fm$^2$~\cite{1989-GSI}. Indeed, the authors of Ref.~\cite{1989-GSI} observed that when $F_{548}^\mathrm{SE}$ was combined with $\delta r^2_{109,107}$ from muonic atoms, a surprising crossing of isotopic chains appeared around $Z=50$. To remedy this issue, they elected to interpolate $\delta r^2_{109,107}$ from that of neighbouring isotones.
These epicycles resulted in $K_{548}^\mathrm{SE}$$=4.4(2.7)~$THz~u which was considered to agree with the HF calculation by Bauche~\cite{1974-HF}.
However, as seen in Table~\ref{tab:factors}, a HF calculation can not reproduce the sign of the SMS, which value dominates that of the total MS.
Considerably reducing the errors of the data-points in Fig.~\ref{fig:Fit}, or introducing new ones via measurements, would help shed light on these issues while reducing uncertainties in the extracted radii.

Our recommended $\delta r^2_{109,A}$ are compared with prior extractions in Table~\ref{tab:rad} and Fig.~\ref{fig:rad}. 
Although significantly different IS factors are used, we find agreement with the radii given in Ref.~\cite{2021-Ag} within uncertainties. This is due to the mitigating effect of enforcing $\delta r^2_{109,107}$ from muonic atoms.
However, the all-optical radii have smaller uncertainties, by up to a factor of 5 (for $\delta r^2_{109,105}$).
This motivates more accurate IS measurements for the neutron deficient Ag, whose radii uncertainties are now dominated by experiment.
A larger disagreement is observed when comparing our results with those of the GSI group~\cite{1989-GSI}, with the largest deviation ($4\sigma$) for $\delta r^2_{109,106m}$.
This is due to the reasons described above; namely the different estimation of $\delta r^2_{109,107}$ (see Table~\ref{tab:Ritz}), and the highly different $F_{548}$. 
Whereas most of the nuclei whose ISs were measured with the $548\,$nm line at GSI were also measured with the cooling line later, $^{105}$Ag, $^{105m}$Ag, and $^{106m}$Ag were only measured with the $548\,$nm line. This work endows them with reliable and precise $\delta r^2_{109,A}$.

We conclude this section by comparing $\delta r^2_{109,A}$ from this work, and the value calculated by state of the art density functional theory (DFT) as done in~\cite{2021-Ag} and shown in Fig.~\ref{fig:rad}.
Focusing first on the neutron rich isotopes, we see agreement for the odd-odd nuclei, and a disagreement for the odd-even ones.
On the proton-rich side, an agreement is seen for all nuclei except for $^{96}$Ag, as discussed in~\cite{2021-Ag}, and $^{102}$Ag, which lies 6 standard errors from the DFT calculation.
These discontinuities are further emphasized when looking at the ladder-type differences, given in the last column of Table~\ref{tab:rad}.

\begin{table*}[tbp]
\caption{
Extraction of the difference in mean-squared nuclear charge radius, $\delta r^2$ in fm$^2$, between radioactive silver isotopes and isomers, from optical isotope shifts, $\delta \nu$ in GHz.
When two references are given, the value is their weighted average.
Numbers in bold are an input to the King Plot of Fig.~\ref{fig:Fit}. 
$\nu_{109,A}^{328, \mathrm{KP}}$ is the mean and standard deviation of the posterior distribution of the $328\,$nm line isotope shift which is output by the fit.
The $\delta r^2$ are calculated from $\nu_{109,A}^{328}$ and, when available, $\nu_{109,A}^{328, \mathrm{KP}}$, using the isotope shift factors from Table~\ref{tab:factors}, taking into account the corrections given in Table~\ref{tab:EEISA}.
The uncertainties in  parenthesis are tied to the experimental isotope shift measurements, and those in square brackets are the total systematic uncertainties discussed in subsection~\ref{Sys}.
The absolute radius can be obtained by adding $R$($^{109}$Ag)$=4.564(2)\,$fm~\cite{2004-FH} in quadrature.
The last column includes the ladder-type difference for ground state nuclei and the isomer shifts for isomers.
} 
\begin{ruledtabular}
\begin{tabular}{r r l r l r rrr r}
\multicolumn{1}{c}{$A$} & \multicolumn{1}{c}{$\delta\nu_{109,A}^{548}$} & \multicolumn{1}{l}{Ref.} & \multicolumn{1}{c}{$\delta\nu_{109,A}^{328}$} & \multicolumn{1}{l}{Ref.} & \multicolumn{1}{c}{$\delta\nu_{109,A}^{328, \mathrm{KP}}$} & \multicolumn{1}{c}{$\delta r_{109,A}^{2}$} & Ref.~\cite{2021-Ag} & Ref.~\cite{1989-GSI} & \multicolumn{1}{c}{$\delta r_{A,A+2}^{2}$}  \vspace{2mm}\\
96  &            & & $2.64^{+1.1}_{-0.8}$ &\cite{2021-Ag}  & & $-1.26^{+22}_{-31}[4]$ & $-1.21^{+27}_{-19}[14]^\dagger$ && $-0.20(27)[1]$ \\
97  &            & & $4.36(28)$ &\cite{2014-LISOL,2021-Ag} & & $-1.69(8)[5]$ & $-1.55(8)[15]$ && $0.41(8)[1]$ \\
98  &            & & $3.66(21)$ &\cite{2014-LISOL,2021-Ag} & & $-1.45(6)[4]$ & $-1.36(6)[13]$ && $0.31(8)[1]$  \\ 
99  &            & & $3.20(9)$  &\cite{2014-LISOL,2021-Ag} & & $-1.28^{+2}_{-3}[4]$ & $-1.20^{+3}_{-2}[12]^\dagger$ && $0.27(5)[1]$  \\ 
100 &            & & $2.85(20)$ &\cite{2014-LISOL,2021-Ag} & & $-1.14(6)[3]$ & $-1.02(8)[11]$ && $0.43(7)[1]$\\ 
101 & $\mathbf{4.672(5)}$ &\cite{1989-GSI}& $\mathbf{2.54(23)}$ &\cite{2014-LISOL,2021-Ag} & $2.55(17)$ &$-1.02(5)[3]$& $-0.98(10)[10]$ & $-0.670(3)[135]$ & $0.31(5)[1]$\\
102 &            &                & $1.60(17)$ &\cite{2021-Ag} & & $-0.712(48)[20]$ & $-0.67^{+5}_{-4}[8]^\dagger$&& $0.06(6)[0]$\\
103 & $\mathbf{3.302(4)}$ &\cite{1989-GSI}& $\mathbf{1.58(30)}$ &\cite{2021-Ag} & $1.74(8)$  &$-0.711(23)[24]$&$-0.63(7)[7]$ & $-0.482(2)[98]$& $0.313(24)[10]$\\
104 & $\mathbf{2.939(3)}$ &\cite{1989-GSI}& $\mathbf{1.71(22)}$ &\cite{2021-Ag} & $1.65(14)$ &$-0.648(39)[22]$&$-0.61(6)[6]$ & $-0.416(2)[83]$\\
105 & $1.926(6)$  &\cite{1989-GSI}&&& $0.894(22)$ &$-0.397(6)[13]$&& $-0.296(2)[63]$ & $0.193(6)[6]$\\
107 & $\mathbf{0.9781(5)}$ &\cite{2001-548}& $\mathbf{0.4732(7)}$ & TW && $-0.207(0)[6]$ & $-0.198(2)[20]$ & $-0.148(1)[31]$ & $0.207(0)[6]$\\
114 &            & & $-0.850(3)$ &\cite{2021-Ag} & & $0.408(1)[14]$ & $0.384(1)[50]$  &&$0.116(3)[4]$\\ 
115 &            & & $-0.995(5)$ &\cite{2021-Ag} & & $0.480(1)[16]$ & $0.454(3)[60]$  &&$0.114(2)[4]$\\ 
116 &            & & $-1.040(9)$ &\cite{2021-Ag} & & $0.524(3)[18]$ & $0.500(10)[60]$ &&$0.107(3)[4]$\\ 
117 &            & & $-1.181(6)$ &\cite{2021-Ag} & & $0.595(2)[20]$ & $0.568(3)[70]$  &&$0.107(2)[4]$\\ 
118 &            & & $-1.203(5)$ &\cite{2021-Ag} & & $0.631(1)[21]$ & $0.607(3)[80]$  &&$0.108(2)[3]$\\ 
119 &            & & $-1.348(5)$ &\cite{2021-Ag} & & $0.702(1)[23]$ & $0.675(3)[90]$  &&$0.090(2)[4]$\\ 
120 &            & & $-1.379(4)$ &\cite{2021-Ag} & & $0.740(1)[25]$ & $0.715(2)[90]$  \\ 
121 &            & & $-1.461(3)$ &\cite{2021-Ag} & & $0.791(1)[26]$ & $0.767(1)[100]$ \\ 
&&&&&&&&&\multicolumn{1}{c}{$\delta r_{A,Am}^{2}$}\vspace{2mm}\\
$~99m$ &   & & $3.58(68)$ &\cite{2014-LISOL,2021-Ag} & & $-1.39(19)[5]$ & $-1.18^{+26}_{-22}[12]$ &&$-0.11(19)[0]$\\ 
$101m$ &&& $2.40(24)$ &\cite{2014-LISOL} &  &$-0.98(7)[3]$& $-0.91(6)[10]$ &  & $0.04(8)[0]$\\
$105m$ & $2.229(10)$ &\cite{1989-GSI}&&& $1.21(8)$   &$-0.485(21)[16]$&& $-0.321(2)[65]$ &$-0.088(22)[3]$\\
$106m$ & $2.049(30)$ &\cite{1989-GSI}&&& $1.30(18)$  &$-0.474(52)[16]$&& $-0.271(4)[52]$\\
$108m$ &            & & $0.443(9)$& \cite{1975-108} &&$-0.159(3)[5]$&& $-0.120(13)[20]$\\
$110m$ &            & & $-0.689(60)$ &\cite{1975-108}   &&$0.0551(17)[18]$&& $ 0.036(5)[16]$
\end{tabular}
$^\dagger$ Our analysis suggests that when adopting the tabulated asymmetric statistical uncertainties in $\delta\nu_{109,A}^{328}$, the signs of the corresponding uncertainties in $\delta r_{109,A}^{2}$ should be swapped as compared with those given in Table 1 and Figure 2 of Ref.~\cite{2021-Ag}.
\end{ruledtabular}
\label{tab:rad}
\end{table*}

\section{Summary and outlook}
Nuclear charge radius differences in the silver isotopic chain deviate between experiments at the few sigma level, as seen in Fig.~\ref{fig:rad}. 
To find the origin of these deviations, and reconcile them, we performed high accuracy calculations of isotope shift factors in the low-lying states of atomic silver (Tables~\ref{tab:E} and~\ref{tab:factors}), as well as new precise spectroscopic measurements in the silver $5s$$~^2S_{1/2}$$-5p~^2P_{3/2}$ line (Table~\ref{tab:ExpData}), and combined this information in a global analysis (Tables~\ref{tab:Ritz} and~\ref{tab:rad}).

We find that discrepancies in reported values of $\delta r^2$ are largely the result of the field shift factors $F$ used to extract them from experimental data, rather than discrepancies between experiments.
Our measurement of the center-of-gravity isotope shift is within 2 combined standard errors from a recent collinear laser spectroscopy measurement~\cite{2021-Ag}.
This difference is too small to explain the tensions with earlier studies, narrowing down the suspects to the used IS factors and/or the mean-squared radius difference of the stable pair.
To test the latter, we extracted the mean-squared radius difference of the stable pair from the isotope shifts of four transitions to the ground-state, two of which make use of our new measurement in the $328\,$nm line (see Table~\ref{tab:Ritz}). 
We find a reasonable agreement between these radii, 
which allows for estimating the accuracy of the calculated IS factors.
Based on this, we recommended a radius difference for the stable pair.
It is one combined standard error from the one extracted from muonic atom cascade X-ray spectroscopy~\cite{2004-FH}, and two standard errors away from the value used in prior works~\cite{1989-GSI,2014-LISOL}, which was interpolated from neighbouring nuclei.
The three values of the the mean-squared radius differences used by us and in prior works partially explain the disagreements in the silver chain, which must thus originate from the choice of field shift factor.
We show this by making a projection of our calculated factors from the $5s$$~^2S_{1/2}$$-5p~^2P_{3/2}$ line to the $4d^95s^2~^2D_{5/2}-6p~^2P_{3/2}$  line using a King Plot (Fig.~\ref{fig:Fit}), which also benefits from our new precise IS measurement. The projected factors of the $548\,$nm line disagree by four combined standard errors with the ones evaluated semi-empirically, thus pointing that this is the main culprit of the disagreements.

Having shed light on prior disagreements we provide in Table~\ref{tab:rad} transparent and reliable mean-squared radius differences in the silver isotopic chain.
Their trend is found to be generally consistent with that from a state-of-the-art nuclear theory calculation.
Nevertheless, the calculated shape-staggering effect is overestimated on the neutron-rich side, and there are discontinuities around $A=96$ and $A=102$ which call for further attention, as seen in Fig.~\ref{fig:rad}.
With the much smaller systematic uncertainties afforded by this work, the radii of proton-rich silver nuclei could now be greatly improved with more accurate measurements, and the maximum information can be extracted from new measurements extending even further towards the drip lines~\cite{2024-95}.

\newpage
\begin{acknowledgments}
We thank Mikael Reponen for useful comments. B.K.S. acknowledges the use of ParamVikram-1000 HPC facility at Physical Research Laboratory (PRL), Ahmedabad for carrying out the atomic calculations and his work at PRL is supported by the Department of Space, Government of India. B.O. is thankful for the support of the Council for Higher Education Program for Hiring Outstanding Faculty Members in Quantum Science and Technology.
\end{acknowledgments} 

\bibliography{references}

\end{document}